\documentclass[reprint,aps,prl,twocolumn,groupedaddress,nobibnotes]{revtex4-1} 

\usepackage{graphicx}  
\graphicspath{{../pdf/}{C:/Users/Figures}}
\usepackage{dcolumn}   
\usepackage{bm}        
\usepackage{amssymb}   
\usepackage{amsmath}   
\usepackage{braket}    
\usepackage[caption=false]{subfig}
\usepackage{hyperref}
\usepackage{color}
\usepackage{ulem}
\begin{document}

\title{Phase effects in coherently-stimulated down-conversion with a quantized pump field}
\author{
	Richard J. Birrittella$^{*}$, Paul M. Alsing$^{*}$, and Christopher C. Gerry$^{\dagger}$\\
	\textit{$^{\dagger}$Department of Physics and Astronomy, Lehman College,\\
		The City University of New York, Bronx, New York, 10468-1589,USA}\\
	\textit{$^{*}$Air Force Research Lab, Information Directorate, Rome, NY, USA\\}
}

\date{\today}

\begin{abstract}
	We investigate the effect of the cumulative phase on the photon statistics of the three-mode state whose evolution is described by the trilinear Hamiltonian $\hat{H}_{I}=i\hbar\kappa\big(\hat{a}\hat{b}\hat{c}^{\dagger}-\hat{a}^{\dagger}\hat{b}^{\dagger}c\big)$, wherein the pump is taken to be quantized (and prepared in a coherent state) and the signal and idler modes are initially seeded with coherent states.  We provide a brief review of the two-mode squeezed coherent states generated by non-degenerate coherently-stimulated parametric down-conversion, whereby the nonlinear crystal is driven by a strong classical field.  The statistics of the resulting two mode state have been shown to depend greatly on the cumulative phase $\Phi=\theta_{s}+\theta_{i}-2\phi$ where $\theta_{s\left(i\right)}$ are the signal(idler) coherent state phases and $2\phi$ is the classical pump phase.  Using perturbation theory, we analytically show for short times how the photon statistics and entanglement properties of the resultant state depends strictly on this phase combination.  We also present numerical results  of the relevant quantities to show the evolution of the three modes and provide a qualitative analysis of the steady state valid for long times.  
\end{abstract}

\maketitle

\section{\label{sec:level1}I. Introduction}

Parametric down-conversion has been a readily available source of non-classical light  for many years \cite{ref:hillery}, and has seen applications in the fields of quantum information processing \cite{ref:qit}, quantum metrology \cite{ref:quantMet} and quantum imaging \cite{ref:quantImg}.  Most often considered is the light generated through the process of spontaneous down-conversion, which produces the well known two-mode squeezed vacuum state (TMSVS) \cite{ref:GerryBook}. In this case a strong classical UV field is used to drive a nonlinear crystal producing pairs of frequency down-converted (infrared) photons in the signal and idler modes, which are initially prepared in vacuum states. Recently, Birrittella \textit{et al.} \cite{ref:1} studied the use of coherently-stimulated parametric down-conversion in the context of quantum optical interferometry, wherein the signal/idler modes are initially seeded with coherent states, and noted the interesting effects of the cumulative phase of the two coherent light fields and the pump field, the latter field assumed to be a classically prescribed field. One can define the cumulative phase from the mixing of the three fields as $\Phi = \theta_{s} + \theta_{i}-2\phi$, where $\theta_{s}\;\text{and}\;\theta_{i}$ are the seeded signal and idler coherent state phases, respectively, and $2\phi$ is the phase of the pump field. This cumulative phase was identified in \cite{ref:3} as the Gouy Phase \cite{ref:Gouy}.  In previous studies of these states, such as the one by Caves \textit{et al.} \cite{ref:2}, the phases are all set to zero with the claim that there is no loss of generality by doing so. But this is not the case as it was found that the photon statistics of the state greatly depends on the value of this cumulative phase $\Phi$. This phase manifests more clearly depending on the ordering of the displacement and squeezing operators when defining the state. In the literature, the two-mode squeezed coherent states (TMSCS) are mathematically defined in two ways having to do with the orderings of these operators acting on the double vacuum state.  The states generated are mathematically equivalent but differ in their implied methods of physical generation. From an experimental point of view, we believe that the natural way to think about the states is to assume coherent light beams are fed into the input signal and idler modes of the down-converter which then acts to squeeze those input states—hence the states are the result of coherently-stimulated down-conversion. As the coherent states may be defined as displaced vacuum states, it follows that the TMSCS is mathematically defined by the action of the displacement operators on the vacuum states of each mode of each followed by the action of the two-mode squeeze operator. However, in the literature, specifically the papers of Caves \textit{et al.} \cite{ref:2} and Selvadoray \textit{et al.} \cite{ref:3}, one finds a definition of the TMSCS with the operators acting in reverse order, i.e. with two-mode squeeze operator acting on the double vacuum followed by the displacement operator such that the states generated could be called two-mode displaced squeezed vacuum states (TMDSVS). Once again, the definitions are mathematically equivalent with properly chosen displacement parameters, but physically the latter states are generated by performing independent displacements on both modes of the two-mode squeezed vacuum. That does not appear to be an attractive method for generating the states in the laboratory in view of the fact that displaced vacuum states (coherent states) are readily available from well phase-stabilized lasers. As we show in Section II, this more “natural” ordering of displacement followed by two-mode squeezing more clearly displays the effects of the cumulative phase. 

For the method of generation we had in mind, that is, where the two displacement operators act on the double vacuum followed by the action of the two-mode squeezing operation, the average photon number in the quantized output signal and idler beams strongly depends on the combined phase angle $\Phi$.  In the reverse ordering of the operators where the two-mode squeezing operator acts on the double vacuum states and where the two displacement operators act on each of the modes of the generated squeezed vacuum states, the phase effects are still there but are somewhat hidden, as pointed out in \cite{ref:1}.  Furthermore, the joint photon number distributions for the output signal and idler beams are remarkably altered by changing $\Phi$ from $0$ to $\pi$, with the value of $\pi$ yielding the greatest average photon number for a given set of parameters.  Effectively, the phases of the input fields combined such that $\Phi=\pi$ strongly amplify the signal and idler beams. On the other hands, if the output signal and idler beams are subject to arbitrary unitary transformation, the individual phases may have some affect. As a specific example, it was shown \cite{ref:1} that having the signal and idler modes incident on a 50:50 beam splitter will result in a joint-photon number distribution that depends heavily on the choice of individual phases.      

The work described in \cite{ref:1}, as indicated above, is performed within the confines of the parametric approximation in which the pump field is treated as a classical undepleted field.  Work has been performed in considering pump depletion with seeded vacuum states \cite{ref:4} with an emphasis on modeling black hole dynamics. The authors of \cite{ref:4} on to show entanglement between late-time Hawking radiation and the black hole's quantum gravitational degrees of freedom. This work was expanded on in \cite{ref:5} and \cite{ref:6}.  We expand on this work, within the framework of quantum optics and quantum state engineering, by considering the questions: What is the effect of the quantization of the pump field on the cumulative phase effects seen when the signal and idler modes are initially seeded with coherent states?  How will the state statistics and entanglement properties of the resultant three mode state evolve in time as the cumulative and individual phases, determined by the initially prepared three mode state, are varied?

The paper is organized as follows: In Section II we briefly review the two-mode squeezed states and their production by spontaneous down-conversion. The effects of the choices of the phases of the two input coherent states and of the classical pump field in combination, referred to as the cumulative phase, are studied as a means of controlling the properties of the output fields. In Section III we first employ a perturbative approach wherein the pump, signal, and idler fields are all quantized and initially prepared in coherent states in order to analytically describe how the state evolves for short times. In Section IV, we present results based on the direct numerical integration of the relevant coupled differential equations to obtain non-perturbative results, subject to the same initial conditions. In all cases, we emphasis the effects of the cumulative phase on the state statistics as well as the entanglement properties of the system.  In Section V, we qualitatively discuss the long-time steady state. Lastly, in Section VI we conclude with a brief summary and some closing remarks.  

\section{\label{sec:level2} II. Phase Effects in Coherently Stimulated Down-Conversion in the Parametric Approximation}

As is well known, non-degenerate parametric down-conversion has been a reliable source of two-mode non-classical states of lights in the laboratory for years \cite{ref:hillery}.  In the parametric approximation wherein the pump field is assumed undepleted, the interaction Hamiltonian for the down-conversion process is given by \cite{ref:7}
\begin{equation}
	\hat{H}_{\text{I}} = i\hbar\left(\gamma\hat{a}\hat{b} - \gamma^{*}\hat{a}^{\dagger}\hat{b}^{\dagger}\right).
	\label{eqn:7}
\end{equation}

\noindent The parameter $\gamma$ is proportional to the second order nonlinear susceptibility $\chi^{\left(2\right)}$ and to the amplitude and phase factor of the pump laser field, assumed here to be a strong classical field such that depletion and fluctuations in the field can be ignored. The quantized field modes $a$ and $b$ are taken to be the signal and idler fields, respectively.  The two-mode squeeze operator $\hat{S}\left(z\right)$ is realized as \cite{ref:GerryBook}
 
\begin{equation}
	\hat{S}\left(z\right) = e^{-i\hat{H}_{\text{I}}t/\hbar} = e^{r\left(\hat{a}\hat{b}e^{-2i\phi}- \hat{a}^{\dagger}\hat{b}^{\dagger}e^{2i\phi}\right)},
	\label{eqn:1}
\end{equation}

\noindent where we have written $\gamma=|\gamma|e^{2i\phi}$ and where the squeeze parameter $r=|\gamma|t$ can be understood as a scaled dimensionless time.  Typically the signal and idler beams are initially in vacuum states, and thus the output state will be the two-mode squeezed vacuum state (TMSVS) $\ket{\xi}$ given by

\begin{align}
	\ket{\xi} &= \hat{S}\left(z\right)\ket{0}_{a}\ket{0}_{b} = \left(1-|z|^{2}\right)^{-1/2}\sum_{n=0}^{\infty}z^{n}\ket{n}_{a}\ket{n}_{b}\nonumber \\
	&= \frac{1}{\cosh r}\sum_{n=0}^{\infty}\left(-1\right)^{n}e^{2in\phi}\tanh^{n}r\ket{n}_{a}\ket{n}_{b}.
\label{eqn:4}
\end{align}

\noindent Note that $\gamma$ and $2\phi$ are the amplitude and phase of the classical pump field, respectively.  The total average photon number is given by
	
\begin{widetext}
	\begin{align}
		\bar{n}_{\text{total}} &=\braket{\psi_{\text{in}}|\hat{S}^{\dagger}\left(z\right)\left(\hat{a}^{\dagger}\hat{a} + \hat{b}^{\dagger}\hat{b}\right)\hat{S}\left(z\right)|\psi_{\text{in}}} \nonumber \\
		&= \braket{\psi_{\text{in}}|\bigg[\left(\hat{a}^{\dagger}\hat{a} + \hat{b}^{\dagger}\hat{b}\right)\cosh 2r -  \left(e^{2i\phi}\hat{a}^{\dagger}\hat{b}^{\dagger}+e^{-2i\phi}\hat{a}\hat{b}\right)\sinh 2r + 2\sinh^{2}r\bigg]|\psi_{\text{in}}}
		\label{eqn:2}
	\end{align}
\end{widetext}

\noindent where we have used the operator relations

\begin{equation}
	\hat{S}^{\dagger}\left(z\right)
		\begin{pmatrix}
			\hat{a}  \\
			\hat{b} 
		\end{pmatrix} 
	\hat{S}\left(z\right)
	=
		\begin{pmatrix}
			\hat{a}\cosh r - e^{2i\phi}\hat{b}^{\dagger}\sinh r  \\
			\hat{b}\cosh r - e^{2i\phi}\hat{a}^{\dagger}\sinh r 
		\end{pmatrix},
	\label{eqn:3} 
\end{equation}

\noindent obtained by use of the Baker-Hausdorff lemma \cite{ref:BCH}.  For the case of an input double vacuum state, $\ket{\psi_{\text{in}}} = \ket{0}_{a}\ket{0}_{b}$, the total average photon number is that of the squeezed vacuum state, given by $\bar{n}=2\sinh^{2}r$. Here the average photon number is independent of the pump phase $2\phi$.  The photon states of each mode are tightly correlated and the state as a whole is highly non-classical due to the presence of squeezing in one of the two-mode quadrature operators. The joint photon number probability distribution for there being $n_{1}$ photons in the $a$-mode and $n_{2}$ photons in the $b$-mode is

\begin{equation}
	P\left(n_{1},n_{2}\right) = \big|\braket{n_{1},n_{2}|\xi}\big|^{2} = \frac{\tanh^{2n}r}{\cosh^{2}r}\times\delta_{n_{1},n}\delta_{n_{2},n},
	\label{eqn:5}
\end{equation}

\noindent such that only the diagonal elements $n_{1}=n_{2}=n$ are nonzero.  The photon-number statistics are super-Poissonian in each mode.  In fact, tracing over either mode yields a single mode mixed state with a thermal distribution \cite{ref:therm1} \cite{ref:therm2}.

We now turn to a discussion of the two-mode squeezed coherent states (TMSCS) which we take to be the output state for an input product of coherent states, i.e. $\ket{\psi_{\text{in}}} = \ket{\alpha_{s}}_{a}\otimes\ket{\alpha_{i}}_{b}$ where

\begin{equation}
	\ket{\alpha}=\hat{D}\left(\alpha\right)\ket{0} = e^{-\tfrac{1}{2}|\alpha|^{2}}\sum_{n=0}^{\infty}\frac{\alpha^{n}}{\sqrt{n!}}\ket{n},
	\label{eqn:10}
\end{equation}

\noindent and where $\hat{D}\left(\alpha\right)=e^{\alpha\hat{a}^{\dagger}-\alpha^{*}\hat{a}}$ is the usual displacement operator. The output state, using the compact notation $\hat{D}\left(\alpha_{s},\alpha_{i}\right)=\hat{D}_{a}\left(\alpha_{s}\right)\hat{D}_{b}\left(\alpha_{i}\right)$, is then given by 

\begin{equation}
	\ket{z;\alpha_{s},\alpha_{i}} =\hat{S}\left(z\right)\ket{\psi_{\text{in}}}= \hat{S}\left(z\right)\hat{D}\left(\alpha_{s},\alpha_{i}\right)\ket{0}_{a}\ket{0}_{b}.
	\label{eqn:9}
\end{equation}

\noindent The process generating the state of Eq. \ref{eqn:9} is called, for obvious reasons, coherently-stimulated down-conversion.  The average total photon number for the state given by Eq. \ref{eqn:9} is

\begin{widetext}
	\begin{align}
		\bar{n}_{\text{total}} &=  \braket{\alpha_{s},\alpha_{i}|\bigg[\big(\hat{a}^{\dagger}\hat{a} + \hat{b}^{\dagger}\hat{b}\big)\cosh 2r - \big(\hat{a}\hat{b}e^{-2i\phi} + \hat{a}^{\dagger}\hat{b}^{\dagger}e^{2i\phi}\big)\sinh 2r + 2\sinh^{2}r\bigg]|\alpha_{s},\alpha_{i}}\nonumber \\
		&= \big(|\alpha_{s}|^{2} + |\alpha_{i}|^{2}\big)\cosh 2r + \left(e^{2i\phi}\alpha_{s}^{*}\alpha_{i}^{*} + e^{-2i\phi}\alpha_{s}\alpha_{i}\right)\sinh 2r + 2\sinh^{2} r,
		\label{eqn:12}
	\end{align}
\end{widetext}

\noindent where we have used the results of Eq.\ref{eqn:2} and that

\begin{equation}
	\hat{D}^{\dagger}\left(\lambda\right)\hat{a}\hat{D}\left(\lambda\right)=\hat{a}+\lambda,\;\;\hat{D}^{\dagger}\left(\lambda\right)\hat{a}^{\dagger}\hat{D}\left(\lambda\right)=\hat{a}^{\dagger}+\lambda^{*}.
	\label{eqn:13}
\end{equation}

\noindent Setting $\alpha_{s} = |\alpha_{s}|e^{i\theta_{s}}$ and $\alpha_{i}=|\alpha_{i}|e^{i\theta_{i}}$, we have

\begin{align}
	\bar{n} = \big(|&\alpha_{s}|^{2} + |\alpha_{i}|^{2}\big)\cosh 2r - \nonumber \\
	&-2|\alpha_{s}||\alpha_{i}|\cos\Phi\sinh 2r + 2\sinh^{2}r,
	\label{eqn:14}
\end{align}

\noindent where $\Phi=\theta_{s}+ \theta_{i} - 2\phi$ is the cumulative phase of the interaction. Evidently, the average photon number for the TMSCS depends on the combination of the phases $\theta_{s},\;\theta_{i}\;\text{and}\;2\phi$.  As far as we are aware, the effects of the phases on the average photon number in coherently-stimulated parametric down-conversion, as given in Eq.\ref{eqn:14}, has yet to be demonstrated experimentally.  The joint-photon number distribution also depends only on the value of the $\Phi$ for a given set of parameters.  However, the joint photon-number distribution obtained \textit{after} the two beams are mixed at a 50:50 beam splitter depends on the individual values of the coherent state phases as well as the cumulative phase \cite{ref:1}.  

In the literature, one quite often finds the TMSCS defined according to the reverse ordering of the squeeze and displacement operators acting on the vacuum than was used above.  That is, one finds the alternative representation

\begin{equation}
	\ket{\beta_{s},\beta_{i};z} = \hat{D}\left(\beta_{s},\beta_{i}\right)\hat{S}\left(z\right)\ket{0}_{a}\ket{0}_{b},
	\label{eqn:15}
\end{equation}

\noindent where $\beta_{s}=|\beta_{s}|e^{i\psi_{s}}$ and $\beta_{i}=|\beta_{i}|e^{i\psi_{i}}$ are not the same coherent state amplitudes and phases that appear in Eq. \ref{eqn:9}.  For this ordering, the average total photon number is given by 

\begin{widetext}
	\begin{align}
		\bar{n}_{\text{total}} &= \braket{\beta_{s},\beta_{i};z|\big(\hat{a}^{\dagger}\hat{a}+\hat{b}^{\dagger}\hat{b}\big)|\beta_{s},\beta_{i};z}\nonumber \\
		&= \braket{0,0|\hat{S}^{\dagger}\left(z\right)\hat{D}^{\dagger}\left(\beta_{s},\beta_{i}\right)\big(\hat{a}^{\dagger}\hat{a}+\hat{b}^{\dagger}\hat{b}\big)\hat{S}\left(z\right)\hat{D}\left(\beta_{s},\beta_{i}\right)|0,0}\nonumber \\
		&= |\beta_{s}|^{2} + |\beta_{i}|^{2} +2\sinh^{2}r.
		\label{eqn:16}
	\end{align}
\end{widetext}

\noindent Note that the result displays no obvious dependence on the phases $\psi_{s},\;\psi_{i},\;\text{and}\;2\phi$.  However, the two representations are equivalent provided

\begin{equation}
	\hat{S}\left(z\right)\hat{D}\left(\alpha_{s},\alpha_{i}\right)\hat{S}^{\dagger}\left(z\right) = \hat{D}\left(\beta_{s},\beta_{i}\right),
	\label{eqn:17}
\end{equation}

\noindent which holds true if 

\begin{equation}
	\beta_{s}=\mu\alpha_{s}-\nu\alpha_{i}^{*},\;\;\beta_{i}=\mu\alpha_{i}-\nu\alpha_{s}^{*},
	\label{eqn:18}
\end{equation}

\noindent where $\mu=\cosh r\;\text{and}\;\nu=e^{2i\phi}\sinh r$.  The inverse transformations are

\begin{equation}
	\alpha_{s}=\mu\beta_{s}+\nu\beta_{i}^{*},\;\;\alpha_{i}=\mu\beta_{i}+\nu\beta_{s}^{*},
	\label{eqn:19}
\end{equation}

\noindent so that under these conditions $\ket{z;\alpha_{s},\alpha_{i}}$ and $\ket{\beta_{s},\beta_{i};z}$ are identical states representing different methods of generation with different displacement operator parameters for the different orderings.  

As mentioned, our result for the average photon number calculated when performing the squeezing operator prior to the displacement operation, $\ket{\beta_{s},\beta_{i};z}$ is independent of the phases $\psi_{s},\;\psi_{i},\;\text{and}\;2\phi$.  That is, there is no \textit{explicit} phase dependence.  However, because of the transformations of Eq. \ref{eqn:18} and \ref{eqn:19}, there is an \textit{implicit} dependence on the phases $\theta_{s},\theta_{i},\;\text{and}\;2\phi$ which show up in the cumulative phase $\Phi=\theta_{s}+\theta_{i}-2\phi$ in Eq. \ref{eqn:14}.  In this sense, the phase dependence of Eq. \ref{eqn:14} is "hidden."  Caves \textit{et al.} \cite{ref:2} and Selvadory \textit{et al.} \cite{ref:3} use the definition of Eq. \ref{eqn:15} for the TMSCS, though the latter authors, for calculational convenience, also use the definition given by Eq. \ref{eqn:9}.  Our result in Eq. \ref{eqn:16} agrees with that of Selvadoray \textit{et al.} \cite{ref:3} who point out that $\bar{n}$ is insensitive to a certain a certain combination of angles that here we shall call $\Psi$, which in our notation has the form $\Psi=\psi_{s}+\psi_{i}-2\phi$. 

In the Appendix of \cite{ref:1}, the relationship between the phase angles $\Phi$ and $\Psi$ as well as the relationship between the sets of displacement phases $\left(\theta_{s},\theta_{i}\right)$ and $\left(\psi_{s},\psi_{i}\right)$ is explicitly shown.  The result of Eq. \ref{eqn:14} is not inconsistent with the result of Eq. \ref{eqn:16} as long as the relations of Eqs. \ref{eqn:18} and \ref{eqn:19} hold.  The essential point here is that the phases of the pump field and of the input coherent states, through $\Phi$, can be adjusted so as to exert control over the average photon number of the output field of the down-converter and of the statistics of this field.  For the choices $|\alpha_{s}|=|\alpha_{i}|=|\alpha|$, we have

\begin{equation}
	\bar{n}=2|\alpha|^{2}\big[\cosh 2r - \cos\Phi\sinh 2r\big] + 2\sinh^{2}r.
	\label{eqn:20}
\end{equation}

\noindent For $\Phi=0$, the term within the brackets goes to zero for sufficiently high values of $r$.  We are then left with the dominant contribution $\bar{n}=2\sinh^{2}r$, which is simply the average photon number for the two mode squeezed vacuum state.  Obviously we can maximize $\bar{n}$ for the choice $\Phi=\pi$.  As we show below, these difference choices of $\Phi$ dramatically affect the nature of the photon number distributions both before and after beam splitting.  We also point out that for a fixed value of $\Phi$, different arrangements of the individual phases affect the photon number distribution after beam splitting but not before.  We note that Caves \textit{et al.} \cite{ref:2}, who examined the TMSCS as defined through Eq. \ref{eqn:15}, set $\phi=0$, stating this can be done without loss of generality.  This is misleading as should be clear from the preceding discussion.  

We now proceed to write down the quantum amplitudes associated with the states $\ket{z;\alpha_{s},\alpha_{i}}$.  In terms of the number states,

\begin{equation}
	\ket{z;\alpha_{s},\alpha_{i}} = \sum_{n_{1}=0}^{\infty}\sum_{n_{2}=0}^{\infty}c\left(n_{1},n_{2}\right)\ket{n_{1}}_{a}\ket{n_{2}}_{b}.
	\label{eqn:21}
\end{equation}

\noindent At this point it is useful to convert our two-mode number state labeling to the angular momentum states $\ket{j,m}$ such that we have the mapping \cite{ref:Yurke}

\begin{equation}
	\ket{n_{1}}_{a}\otimes\ket{n_{2}}_{b} = \ket{j,m},\;\;\;j=\frac{n_{1}+n_{2}}{2},\;m=\frac{n_{1}-n_{2}}{2}.
	\label{eqn:22}
\end{equation}

\noindent We can rewrite our state in terms of the angular momentum states as

\begin{equation}
	\ket{z;\alpha_{s},\alpha_{i}} = \sum_{j=0,\tfrac{1}{2},1,..}^{\infty}\sum_{m=-j}^{j}c\left(j+m,j-m\right)\ket{j,m}.
	\label{eqn:23}
\end{equation}

\noindent The state coefficients $c\left(j+m,j-m\right)$, adapted and corrected from a result obtained by Selvadoray \textit{et al.} \cite{ref:3}, are given by

\begin{widetext}
	\begin{align}
		c\left(j+m,j-m\right) = & \; \text{exp}\left[-i\pi\left(j-|m|\right)\right]\bigg(\frac{\left(j-|m|\right)!}{\left(j+|m|\right)!}\bigg)^{1/2}\bigg(\frac{\alpha_{s}\alpha_{i}}{\mu\nu}\bigg)^{|m|}\frac{1}{\mu}\bigg(\frac{\nu}{\mu}\bigg)^{j}\times \nonumber \\
		&\times L_{j-|m|}^{2|m|}\bigg(\frac{\alpha_{s}\alpha_{i}}{\mu\nu}\bigg)\bigg(\frac{\alpha_{s}}{\alpha_{i}}\bigg)^{m}\text{exp}\big[-\tfrac{1}{2}\left(|\alpha_{s}|^{2} + |\alpha_{i}|^{2}\right)\big]\text{exp}\bigg[\frac{\nu^{*}\alpha_{s}\alpha_{i}}{\mu}\bigg],
		\label{eqn:24}
	\end{align}
\end{widetext}

\noindent where once again $\mu=\cosh r$ and $\nu=e^{2i\phi}\sinh r$.  The functions $L_{n}^{k}\left(x\right)$ are the associated Laguerre polynomials.  In terms of the phases $\theta_{1},\;\theta_{2}\;\text{and}\;\phi$ the amplitudes of Eq. \ref{eqn:24} can be written as
\begin{widetext}
	\begin{align}
		c\left(j+m,j-m\right) = & \; \text{exp}\left[-i\pi\left(j-|m|\right)\right]\bigg(\frac{\left(j-|m|\right)!}{\left(j+|m|\right)!}\bigg)^{1/2}\bigg(\frac{2|\alpha_{s}||\alpha_{i}|}{\sinh 2r}\bigg)^{|m|}\frac{\tanh^{j} r}{\cosh r}\times \nonumber \\
		&\times \text{exp}\bigg[i\left(|m|\Phi + 2j\phi\right)\bigg]L_{j-|m|}^{2|m|}\bigg(\frac{2|\alpha_{s}||\alpha_{i}|}{\sin 2r}\bigg)\bigg(|\frac{\alpha_{s}|}{|\alpha_{i}|}\bigg)^{m}\text{exp}\bigg(im\left(\theta_{1}-\theta_{2}\right)\bigg)\times \nonumber \\
		&\times\text{exp}\big[-\tfrac{1}{2}\left(|\alpha_{s}|^{2} + |\alpha_{i}|^{2}\right)\big]\text{exp}\bigg[e^{i\Phi}|\alpha_{s}||\alpha_{i}|\tanh r\bigg].
		\label{eqn:25}
	\end{align}
\end{widetext}  

\noindent Notice the appearance of the cumulative phase $\Phi=\theta_{s}+\theta_{i}-2\phi$.  As noted above, the average total photon number for the two beams in this representation as well as the joint photon number distribution depends only on $\Phi$ as the individual phases only contribute an overall phase factor.  Before moving on to consider the effects of quantizing the pump field, we include a more thorough analysis of the photon statistics of the state for later comparison; we calculate the Mandel $Q$ parameter \cite{ref:MandelQ}, defined formally as

\begin{equation}
	Q = \frac{\Delta^{2}\hat{n}-\braket{\hat{n}}}{\braket{\hat{n}}},
	\label{eqn:mandel1}
\end{equation}

\noindent such that $Q=0$ denotes Poissonian statistics with a photon number variance equal to its average, $Q<0$ denotes sub-Poissonian statistics, useful for quantifying non-classicality of the state  and $Q>1$ denotes super-Poissonian statistics characterized by a very broad spread in the photon number distribution about its average.  For coherent states, the most classical of quantized field states, $Q=0$, and the photon number distribution is Poissonian.  For the most 'quantum' of quantized field states, a Fock state of definite photon number, the photon number variance is zero and the Mandel parameter is $Q=-1$, providing a lower bound on the Mandel $Q$ parameter.  For the case of a TMSCS, setting $|\alpha_{s}|=|\alpha_{i}|=|\alpha|$, we find

\begin{widetext}
	\begin{equation}
		Q_{s}\left(\Phi\right)=Q_{i}\left(\Phi\right)=2\sinh^{2}r-\frac{\sinh^{4}r}{|\alpha|^{2}\left(\cosh 2r-\sinh 2r\cos\Phi\right)+\sinh^{2}r},
		\label{eqn:mandel2}
	\end{equation}
\end{widetext}

\begin{figure}
	\includegraphics[width=0.96\linewidth,keepaspectratio]{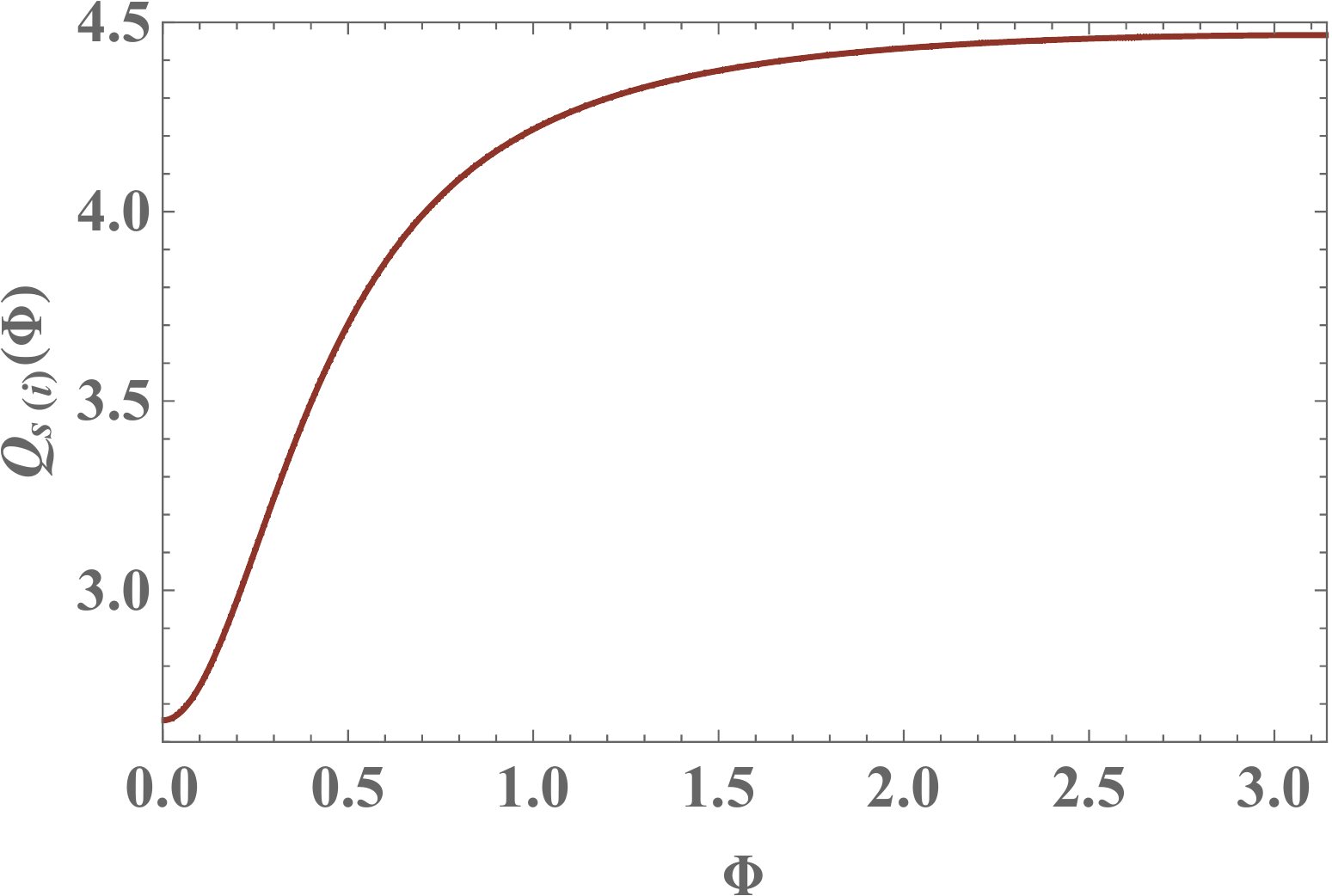}
	\caption{Mandel $Q$ parameter for the signal(idler) mode, with $|\alpha_{s}|=|\alpha_{i}|=|\alpha|=2$ and $r=1.2$ plotted against the cumulative phase $\Phi$.  Both modes remains super-Poissonian for all values of $\Phi$.}
	\label{fig:Mandel}
\end{figure}

\noindent which is plotted Fig.\ref{fig:Mandel} against the cumulative phase $\Phi$. We find that both modes remain super-Poissonian for all values of $\Phi$.  This is consistent with the corresponding joint photon-number distributions detailed in \cite{ref:1} which show the distribution widening, and the peak migrating outwards away from the vacuum, as $\Phi\to\pi$.

We also wish to comment on the entanglement between the signal and idler beams as a function of the cumulative phase.  To this end we will consider two separate quantities: the von Neumann entropy and the logarithmic negativity. For a density operator $\hat{\rho}$, the von Neumann entropy is defined as \cite{ref:vonNeumann}

\begin{equation}
	S\left(\hat{\rho}\right)=-\text{Tr}\left[\hat{\rho}\;\text{ln}\hat{\rho}\right]=-\sum_{k}\rho_{kk}\;\text{ln}\rho_{kk},
	\label{eqn:neumann}
\end{equation}

\noindent where in the last line of Eq. \ref{eqn:neumann} we assume $\hat{\rho}$ has been diagonalized with elements $\rho_{kk}$.  The logarithmic negativity is given by \cite{ref:PPT2} 

\begin{equation}
	E_{\mathcal{N}}\left(\rho_{s,i}\right) =\text{log}_{2}\left[1+2\mathcal{N}\left(\rho_{s,i}\right)\right]= \text{log}_{2}||\rho_{s,i}^{T_{i}}||,
	\label{eqn:log1}
\end{equation} 

\noindent where $\mathcal{N}\left(\rho_{s,i}\right)$ is the 'negativitiy' deriving from the PPT criterion for separability, and serves as an entanglement monontone \cite{ref:PPT}. It is given by  

\begin{equation}
	\mathcal{N}\left(\rho_{s,i}\right) = \frac{||\rho_{s,i}^{T_{i}}||-1}{2} = \sum_{i}|\lambda_{i}^{\left(-\right)}|.
	\label{eqn:log2}
\end{equation}

\noindent where $||\rho_{s,i}^{T_{i}}||$ is the trace norm of the partial transpose with respect to the idler mode of the density operator $\rho_{s,i}$ and where $\lambda_{i}^{\left(-\right)}$ represent the negative eigenvalues of $\rho_{s,i}^{T_{i}}$.   Interestingly enough, for the case of a constant pump, both the von Neumann entropy and the logarithmic negativity are independent of both the choices of the individual phases as well as the cumulative phase. In addition to this, both quantities are independent of the coherent state amplitudes and only depend on the choice of the squeeze parameter $r$.  Furthermore, the logarithmic negativity of the two-mode squeezed coherent states is simply that of the two-mode squeezed vacuum state \cite{ref:PPT2}

\begin{equation}
	E_{\mathcal{N}}\left(\rho\right) = \text{log}_{2}\big(e^{2 r}\big).
	\label{eqn:log3}
\end{equation}

Here we wish to remark on an error made in the analysis found in \cite{ref:1}. The authors remark on the entanglement between signal and idler modes through calculation of the linear entropy, which is the linear approximation of the von Neumann entropy given by  $S_{\text{Lin},s\left(i\right)}=1-\text{Tr}\big[\hat{\rho}^{2}_{s\left(i\right)}\big]$ where $\hat{\rho}_{s\left(i\right)}$ is the reduced density operator of the signal(idler) mode.  They claimed that the linear entropy increases as a function of $\Phi$, that is, the entanglement between signal and idler modes increase as $\Phi\to\pi$.  This is incorrect, as the state was truncated at too small a photon number in the calculations.  In fact, the linear entropy, like the von Neumann entropy, is independent of the cumulative phase.  The two modes of the TMSCS are no more entangled than that of the two-mode squeezed vacuum state for all values of the phase.  

\section{\label{sec:level3} III. Quantizing the pump field - Perturbative Analysis}

So far, we have considered the pump field of the down-converter to be a classically prescribed field, which means we have ignored the effects of pump depletion. Next, we shall study the states produced in the case where the pump field is quantized and initially taken to be a coherent state. Our goal, once again, is to explore the effects of the cumulative phase $\Phi$ on the evolution of the fully quantized model.  More particularly, our interest is in the photon statistics of the output signal, idler and pump modes as well as quantifying the entanglement between the three modes as the state evolves in time. In the fully quantum mechanical model the Hamiltonian that drives the interaction between the three field states is 

\begin{equation}
\hat{H}_{I} = i\hbar\kappa\big(\hat{a}\hat{b}\hat{c}^{\dagger} - \hat{a}^{\dagger}\hat{b}^{\dagger}\hat{c}\big),
\label{eqn:26}
\end{equation}

\noindent where  $\{\hat{a},\hat{a}^{\dagger}\}$ operates on the signal mode, $\{\hat{b},\hat{b}^{\dagger}\}$ the idler mode and $\{\hat{c},\hat{c}^{\dagger}\}$ the pump mode and where the parameter $\kappa$ is a coupling constant proportional to the $\chi^{\left(2\right)}$ nonlinear susceptibility \cite{ref:Kai2}. The trilinear Hamiltonian, Eq. \ref{eqn:26}, has been investigated numerically as early as 1970 both in the context of spontaneous parametric down-conversion \cite{ref:WBref}\cite{ref:8}\cite{ref:9}\cite{ref:10}\cite{ref:11}\cite{ref:12} as well as emission from superradiant Dicke-states \cite{ref:Dicke}\cite{ref:Bonifacio}. The former, \cite{ref:WBref}$-$\cite{ref:12}, investigated the eigenvalues of the tridiagonal matrix representation of Eq. \ref{eqn:26} in the computational 'logical' basis $\ket{n}_{L}$ formed from the three-modes of the down-converter $\ket{n}_{L}=\ket{n_{p_{0}}}_{p}\ket{n}_{s}\ket{n}_{i}$, where $n_{p_{0}}$ are the initial number of photons occupying the pump mode.  The latter, \cite{ref:Dicke}\cite{ref:Bonifacio}, employed the Schwinger realization of the SU(2) Lie algebra \cite{ref:Yurke} to convert the pump-idler modes into the spin-boson representation such that the trilinear Hamiltonian can be written as  $\hat{H}_{I}=i\hbar\kappa\big(\hat{J}^{\left(p,i\right)}_{+}\hat{a}-\hat{J}_{-}^{\left(p,i\right)}\hat{a}\big)$.  They then go on to develop differential-difference equations for the state probability amplitudes $c_{n}=_{L}\braket{n|e^{-i\hat{H}_{I}t/\hbar}|\psi_{\text{in}}}$ of the output state $\ket{\psi}_{\text{out}}=\sum_{n=0}^{\infty}c_{n}\ket{n}_{L}$. The trilinear Hamiltonian can also be expressed in terms of the SU(1,1) Lie Algebra, whereby the signal-idler modes are written in terms of the su(1,1) ladder operators $K_{+}^{\left(s,i\right)}=\hat{a}^{\dagger}\hat{b}^{\dagger}$ and $\hat{K}_{-}^{\left(s,i\right)}=\hat{a}\hat{b}$ \cite{ref:Yurke} such that Eq. \ref{eqn:26} becomes  $\hat{H}_{I}=i\hbar\kappa\big(\hat{c}^{\dagger}\hat{K}_{-}^{\left(s,i\right)}-\hat{c}\hat{K}_{+}^{\left(s,i\right)}\big)$. This was the form of the interaction Hamiltonain considered by Nation and Blencowe \cite{ref:4} and Alsing \cite{ref:5} in their short-time approximation of the state statistics with a pump field taken as an arbitrary pure state.

The time-evolved state resulting from the interaction Hamiltonian Eq. \ref{eqn:26} is given by $\ket{\psi\left(t\right)} = \hat{U}\left(t\right)\ket{\psi\left(0\right)}$, where $\hat{U}\left(t\right) = e^{-i\hat{H}_{I}t/\hbar}$, and where the initial three-mode state is given by  

\begin{equation}
	\ket{\psi\left(0\right)} = \ket{\alpha_{s}}_{s}\otimes\ket{\alpha_{i}}_{i}\otimes\ket{\gamma}_{p},
	\label{eqn:27}
\end{equation}

\noindent where we assume all three modes are initially occupied by coherent states with displacement amplitudes $\alpha_{s}=|\alpha_{s}|e^{i\theta_{s}},\;\alpha_{i}=|\alpha_{i}|e^{i\theta_{i}},\;\text{and}\;\gamma=|\gamma|e^{2i\phi}$. In the balance of this section we take a perturbative approach to analytically show how the state properties evolve at short times with respect to the cumulative phase $\Phi$. 

Working in the Heisenberg picture, we calculate the average photon number in the pump $\bar{n}_{p}$, signal $\bar{n}_{s}$ and idler $\bar{n}_{i}$ by the usual

\begin{widetext}
	\begin{eqnarray}	
		&\bar{n}_{p}\left(t\right) = \braket{\psi\left(t\right)|\hat{c}^{\dagger}\hat{c}|\psi\left(t\right)} = \braket{\psi\left(0\right)|e^{i\hat{H}_{I}t/\hbar}\hat{c}^{\dagger}\hat{c}e^{-i\hat{H}_{I}t/\hbar}|\psi\left(0\right)}, \nonumber \\
		&\bar{n}_{s}\left(t\right) = \braket{\psi\left(t\right)|\hat{a}^{\dagger}\hat{a}|\psi\left(t\right)} = \braket{\psi\left(0\right)|e^{i\hat{H}_{I}t/\hbar}\hat{a}^{\dagger}\hat{a}e^{-i\hat{H}_{I}t/\hbar}|\psi\left(0\right)}, \\
		&\bar{n}_{i}\left(t\right) = \braket{\psi\left(t\right)| \hat{b}^{\dagger}\hat{b}|\psi\left(t\right)} = \braket{\psi\left(0\right)|e^{i\hat{H}_{I}t/\hbar}\hat{b}^{\dagger}\hat{b}e^{-i\hat{H}_{I}t/\hbar}|\psi\left(0\right)}. \nonumber
		\label{eqn:28}
	\end{eqnarray}   
\end{widetext} 

\begin{figure*}
	\centering
	\subfloat[][]{\includegraphics[width=0.32\linewidth,keepaspectratio]{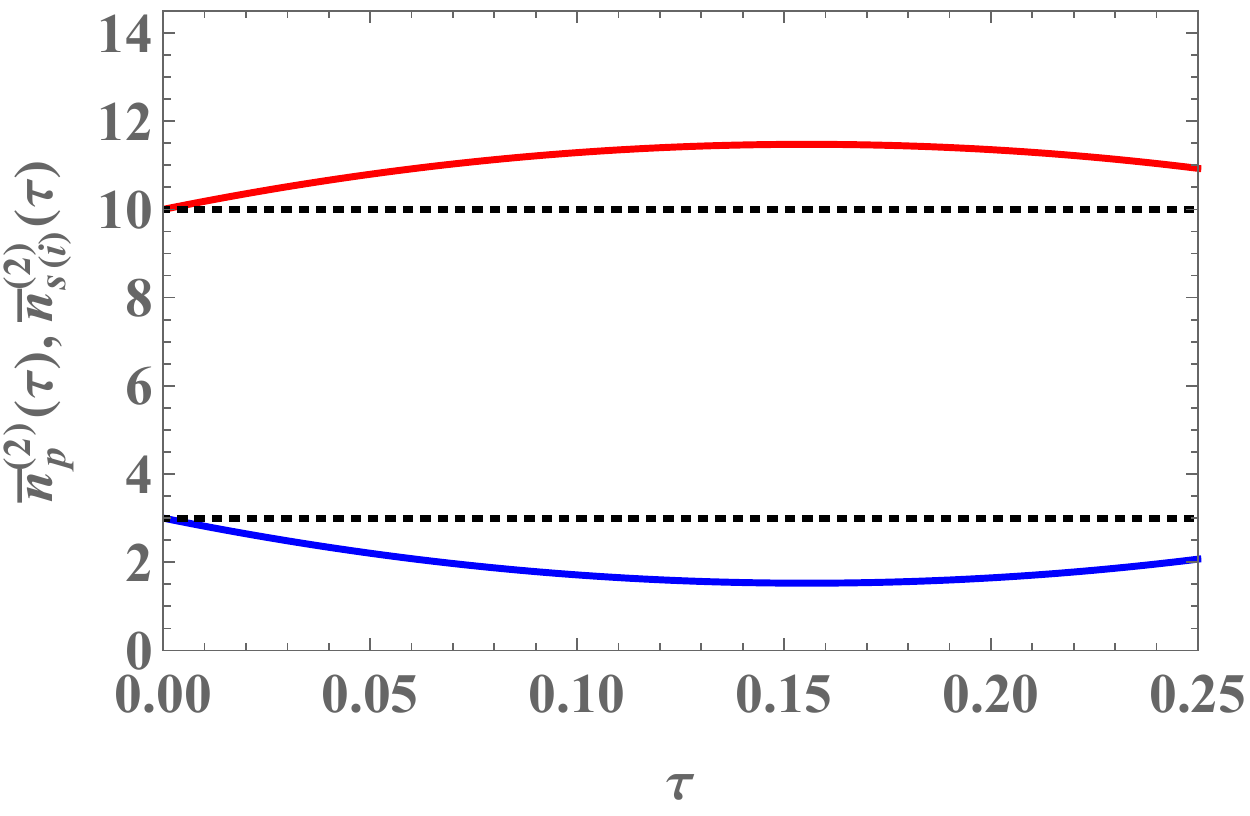}
		\label{fig:AvgPTa}}
	\subfloat[][]{\includegraphics[width=0.32\linewidth,keepaspectratio]{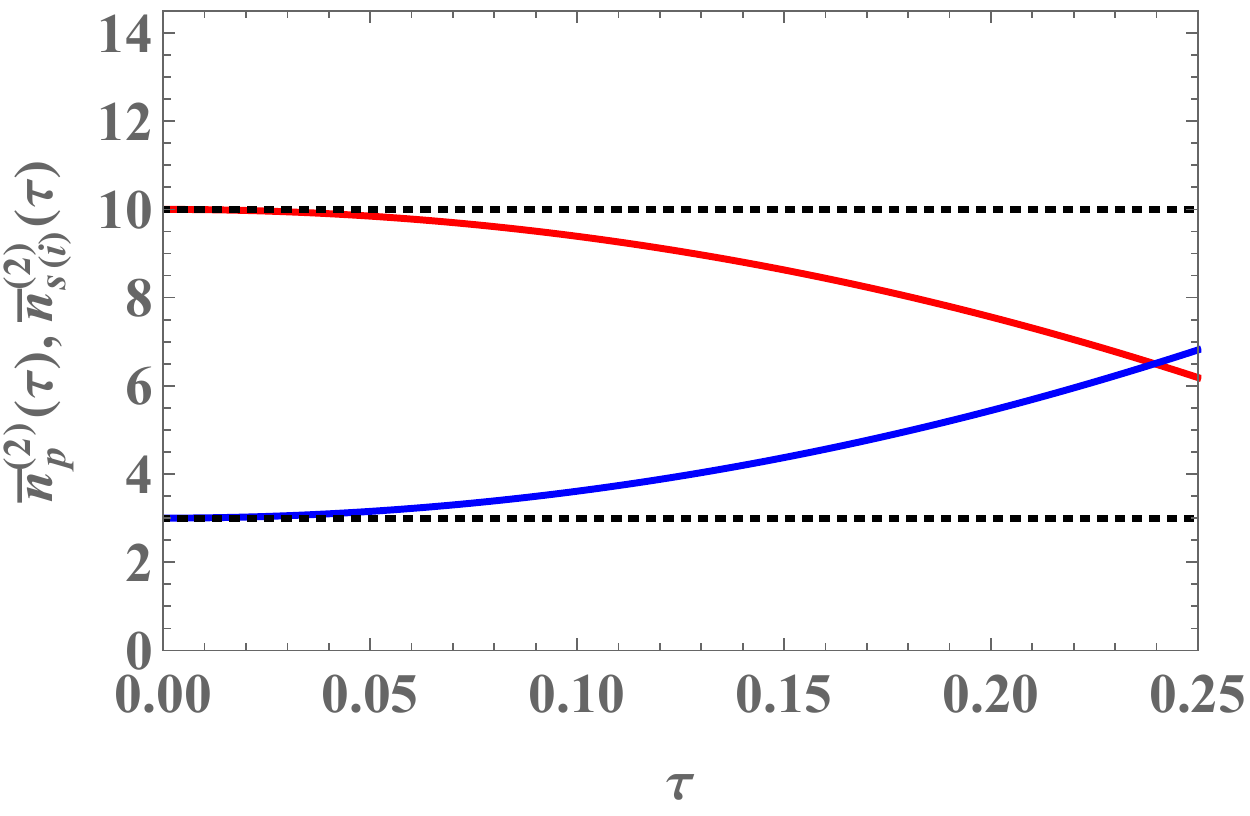}
		\label{fig:AvgPTb}}
	\subfloat[][]{\includegraphics[width=0.32\linewidth,keepaspectratio]{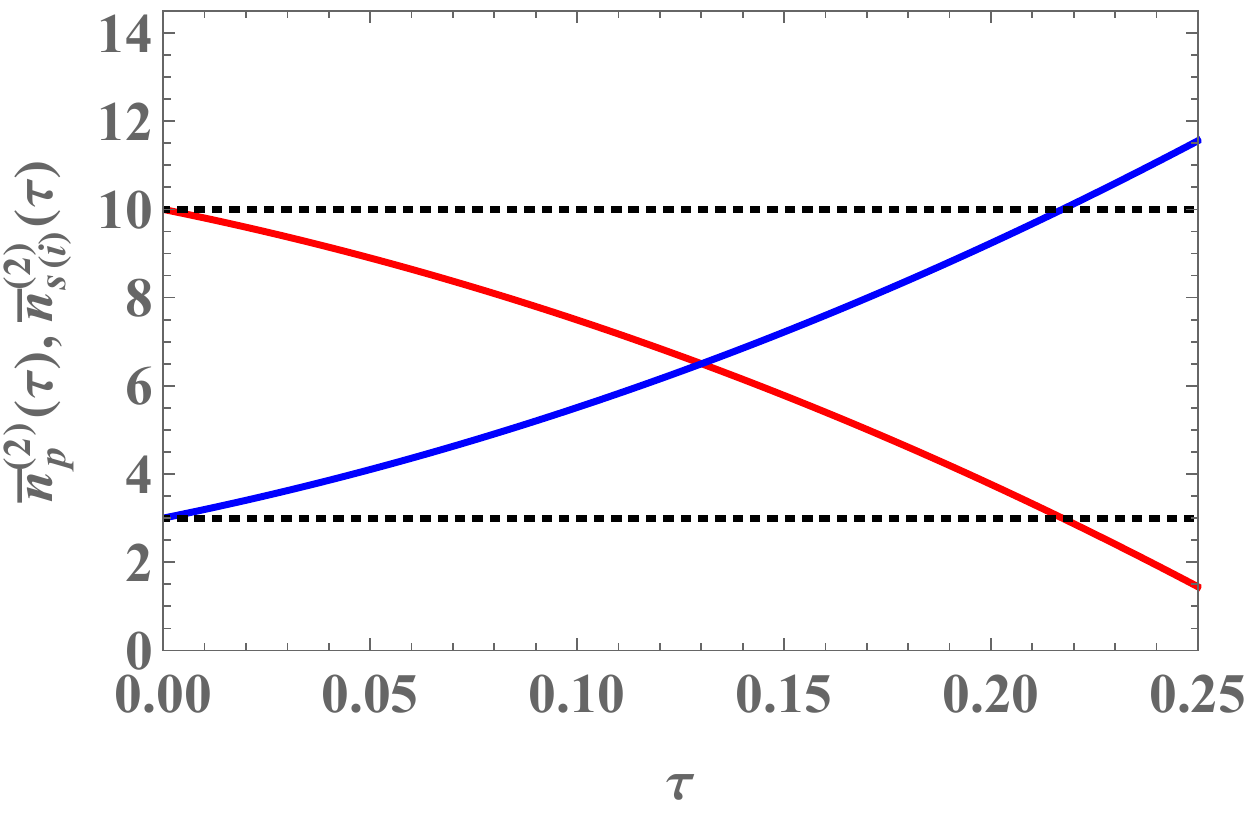}
		\label{fig:AvgPTc}}
	\caption{Average photon number for the pump field (red) and for the signal/idler (blue) plotted against time $\tau$ for \ref{fig:AvgPTa}.) $\Phi=0$, \ref{fig:AvgPTb}.) $\Phi=\pi/2$ and \ref{fig:AvgPTc}.) $\Phi=\pi$. Note that $|\gamma|^{2}=10$ and $|\alpha_{s}|^{2}=|\alpha_{i}|^{2}=3$ for all figures. Dashed black lines denoting the initial pump/signal(idler) average photon numbers are included as a relevant point of reference.}
	\label{fig:AvgPT} 
\end{figure*}

\noindent We can expand the evolution operator in terms of time

\begin{equation}
	e^{\pm i\hat{H}_{I}t/\hbar} = 1 \pm i\hat{H}_{I}t/\hbar - \frac{1}{2!}\big(\hat{H}_{I}t/\hbar\big)^{2} - ....\;.
	\label{eqn:29}
\end{equation}

\noindent  Under this approximation, the average photon number in the pump field is given by 

\begin{equation}
	\bar{n}_{p}\left(\tau\right) = \braket{n_{p}}^{\left(0\right)} + \tau \braket{n_{p}}^{\left(1\right)} + \tau^{2}\braket{n_{p}}^{\left(2\right)}+... \;, 
	\label{eqn:30}
\end{equation}

\noindent where $\tau$ is the scaled dimensionless time $\tau=\kappa t$. Plugging in and calculating each order, we arrive at the result

\begin{align}	
	\braket{n_{p}}^{\left(0\right)} &= |\gamma|^{2},  \nonumber \\ 
	\braket{n_{p}}^{\left(1\right)} &= 2|\alpha_{s}||\alpha_{i}||\gamma|\cos\Phi, \\
	\braket{n_{p}}^{\left(2\right)} &=|\alpha_{s}|^{2}|\alpha_{i}|^{2} - |\gamma|^{2}\big(1+|\alpha_{s}|^{2} + |\alpha_{i}|^{2}\big). \nonumber
	\label{eqn:31}
\end{align}

\noindent It is not surprising when considering the Hamiltonian that drives the interaction, that only odd orders of the average photon number produces a dependency on the cumulative phase $\Phi$. It is also worth pointing out the dependency between each order of the average photon number and the Hamiltonian.  Employing the Baker-Hausdorff lemma \cite{ref:Baker}, the first order correction is given by $\bar{n}_{p}^{\left(1\right)} \propto \braket{\big[\hat{H}_{I},\hat{c}^{\dagger}\hat{c}\big]}$ while the second order correction is $\bar{n}_{p}^{\left(2\right)} \propto \braket{\big[\big[\hat{H}_{I},\hat{c}^{\dagger}\hat{c}\big],\hat{H}_{I}\big]}$, and so forth for higher orders.  In general, for an arbitrary operator $\hat{A}$,

\begin{align}
	\braket{\hat{A}}^{\left(0\right)}&=\braket{\psi\left(0\right)|\hat{A}|\psi\left(0\right)}, \nonumber\\
	\braket{\hat{A}}^{\left(1\right)}&= \frac{i}{\hbar}\braket{\psi\left(0\right)|\big[\hat{H}_{I},\hat{A}\big]|\psi\left(0\right)}, \\
	\braket{\hat{A}}^{\left(2\right)}&= \frac{1}{2\hbar^{2}}\braket{\psi\left(0\right)|\big[\big[\hat{H}_{I},\hat{A}\big],\hat{H}_{I}\big]|\psi\left(0\right)}. 
	\label{eqn:addedin}
\end{align}

\noindent Next we turn our attention towards finding the total average photon number in the signal and idler modes.  Carrying out the same procedure as for the pump field, we can write 

\begin{equation}
	\bar{n}_{s\left(i\right)}\left(\tau\right) = \braket{n_{s\left(i\right)}}^{\left(0\right)} + \tau\braket{n_{s\left(i\right)}}^{\left(1\right)} + \tau^{2}\braket{n_{s\left(i\right)}}^{\left(2\right)}+... \;, 
	\label{eqn:32}
\end{equation}

\noindent where once again we can plug in the approximation made in equation Eq. \ref{eqn:29} and use Eq. \ref{eqn:addedin} to find

\begin{align}	
	\braket{n_{s\left(i\right)}}^{\left(0\right)} &= \;\;|\alpha_{s\left(i\right)}|^{2}, \nonumber \\
	\braket{n_{s\left(i\right)}}^{\left(1\right)} &= -\braket{n_{p}}^{\left(1\right)},  \\
	\braket{n_{s\left(i\right)}}^{\left(2\right)} &=   -\braket{n_{p}}^{\left(2\right)}.
	\label{eqn:3per.8}
\end{align}

\noindent We note that the above could have been found by the Manley-Rowe relations \cite{ref:MRrel} such that

\begin{equation}
	\frac{d\big(\bar{n}_{s}-\bar{n}_{i}\big)}{d\tau}=\frac{d\big(\bar{n}_{s}+\bar{n}_{p}\big)}{d\tau}=\frac{d\big(\bar{n}_{i}+\bar{n}_{p}\big)}{d\tau}=0.
	\label{eqn:manley}
\end{equation}

\noindent The above relations also hold as operator equations as can be seen directly from the Heisenberg equations of motion.  In particular, we note that $d\left[\hat{n}_{p}+\tfrac{1}{2}\left(\hat{n}_{s}+\hat{n}_{i}\right)\right]/d\tau \equiv 0$ which is interpreted as the annihlation of each pump photon creates a (signal/idler) pair of photons.

We plot the average photon numbers for the pump as well as the signal/idler, to second order in $\tau$, in Fig. \ref{fig:AvgPT} against scaled time $\tau$ for $|\alpha_{s}|=|\alpha_{i}|=\sqrt{3}$ and $|\gamma|=\sqrt{10}$. For very short times, where first order perturbation is sufficient, the average photon number for the pump and signal(idler) modes can be expressed 

\begin{align}
	\bar{n}_{p}\left(\tau\right)&=|\gamma|^{2}\Big(1+2\tau\frac{|\alpha_{s}||\alpha_{i}|}{|\gamma|}\cos\Phi\Big), \nonumber\\
	 \\
	\bar{n}_{s\left(i\right)}\left(\tau\right)&=|\alpha_{s\left(i\right)}|^{2}\Big(1-2\tau\frac{|\alpha_{i}||\gamma|}{|\alpha_{s}|}\cos\Phi\Big). \nonumber
	\label{eqn:3per.9}
\end{align}  

\noindent It is readily apparent that for $\Phi=\pi/2$ all three modes remain coherent states. For the choice $\Phi=\pi$ we see a gain in average photon number for the signal/idler modes whilst for $\Phi=0$, back-action from the signal/idler modes result in an initial increase in occupation of the pump mode.     

Next we wish to remark on the photon statistics of the pump and signal/idler modes for short times.  To this end we calculate the second moment of $n$ to first order

\begin{equation}
	\braket{n_{j}^{2}}^{\left(1\right)} = \braket{n_{j}}^{\left(1\right)}\left(1+2\braket{n_{j}}^{\left(0\right)}\right),
	\label{eqn:mandel4}
\end{equation}

\noindent where $j=p,s,i$ for pump, signal and idler, respectively, and to second order 

\begin{widetext}
	\begin{align}
	\braket{n_{p}^{2}}^{\left(2\right)} &= \braket{n_{p}}^{\left(2\right)} + 2|\gamma|^{2}\Big[|\alpha_{s}|^{2}|\alpha_{i}|^{2}\left(2+|\gamma|^{2}\right)-|\gamma|^{2}\left(1+|\alpha_{s}|^{2}+|\alpha_{i}|^{2}+|\alpha_{s}|^{2}|\alpha_{i}|^{2}\right)+|\alpha_{s}|^{2}|\alpha_{i}|^{2}\cos 2\Phi\Big],\nonumber \\
	\\
	\braket{n_{s\left(i\right)}^{2}}^{\left(2\right)} &= \braket{n_{s\left(i\right)}}^{\left(2\right)}+2|\alpha_{s\left(i\right)}|^{2}\Big[|\gamma|^{2}\left(1+|\alpha_{i\left(s\right)}|^{2}\right)\left(2+|\alpha_{s\left(i\right)}|^{2}\right)-|\alpha_{s}|^{2}|\alpha_{i}|^{2}\left(1+|\gamma|^{2}\right)+|\gamma|^{2}|\alpha_{i\left(s\right)}|^{2}\cos 2\Phi\Big]. \nonumber
	\label{eqn:mandel6}
	\end{align}
\end{widetext}

\noindent Higher order corrections of $\braket{n_{j}}\;\text{and}\;\braket{n_{j}^{2}}$ will remain strictly functions of  $\cos\Phi$, that is, $\braket{n_{j}^{2}}^{\left(k\right)}\propto\cos k\Phi+B$, where $B$ is strictly a function of the initial state amplitudes; this can be verified by seeing how the boson operators in the Hamiltonian come in for higher orders. The photon number fluctuations out to second order in all three modes are found to be 

\begin{widetext}
	\begin{align}
		\Delta^{2}n_{p}\left(\tau\right) &=  \braket{n_{p}}^{\left(0\right)} + \tau\braket{n_{p}}^{\left(1\right)}+\tau^{2}\braket{n_{p}}^{\left(2\right)} + .., \\
		\nonumber \\ 
		\Delta^{2}n_{s\left(i\right)}\left(\tau\right) &=  \braket{n_{s\left(i\right)}}^{\left(0\right)} + \tau\braket{n_{s\left(i\right)}}^{\left(1\right)}+\tau^{2}\left(\braket{n_{s\left(i\right)}}^{\left(2\right)}+2|\alpha_{s\left(i\right)}|^{2}|\gamma|^{2}\right) + ..\;.
		\label{eqn:mandel7}
	\end{align}  
\end{widetext}

\noindent We go on to calculate the Mandel $Q$ parameter, Eq. \ref{eqn:mandel1}, in this regime. We point out that the Mandel $Q$ parameter will be strictly a function of the cumulative phase $\Phi$, as only diagonal elements of the density matrix are required to determine the expectation values of the quantities above, and only off-diagonal elements will depend on the individual state phases. The zeroth order term, $Q_{j}^{\left(0\right)}$, is zero as all three modes are initially coherent states. Likewise, it is easy to see using the expectation values above, that to first order in $\tau$ the Mandel $Q$ parameter is zero for all values of the phase $\Phi$ for all modes; for very short times, all three modes remain Poissonian.  To second order, the Mandel $Q$ parameter is found by

\begin{equation}
	Q_{j}^{\left(2\right)}\left(\tau\right)=\frac{\tau^{2}\left[\braket{n_{j}^{2}}^{\left(2\right)}-\braket{n_{j}}^{\left(1\right)\;2}-\braket{n_{j}}^{\left(2\right)}\left(1+2\braket{n_{j}}^{\left(0\right)}\right)\right]}{\braket{n_{j}}^{\left(0\right)}+\tau\braket{n_{j}}^{\left(1\right)}+\tau^{2}\braket{n_{j}}^{\left(2\right)}}.
	\label{eqn:mandel5}
\end{equation}

\noindent Plugging in from the equations above we find that to second order, the Mandel $Q$ parameter remains zero for all values of the phase for the pump mode while the signal/idler modes display decreasingly-growing super-Poissonian statistics as the cumulative phase is varied from $\Phi=0\to\pi$.  

Next we investigate the role of the phases on the entanglement properties of the three-mode state.  Working in the Schr\"{o}dinger picture, we can write the time-dependent state as.. 

\begin{align}
	\ket{\psi\left(\tau\right)}&=\left(1-it\hat{H}_{I}/\hbar+..\right)\ket{\psi\left(0\right)} \nonumber \\
	&=\sum_{\{N\}}\left(C_{N_{p},N_{s},N_{i}}^{\left(0\right)}+\tau C_{N_{p},N_{s},N_{i}}^{\left(1\right)}+...\right)\ket{N_{p},N_{s},N_{i}} \nonumber \\
	&=\sum_{\{N\}}\sum_{i}\tau^{i}C_{N_{p},N_{s},N_{i}}^{\left(i\right)}\ket{N_{p},N_{s},N_{i}}.
	\label{eqn:logneg1}
\end{align}

\noindent To first order in time, the two-mode reduced density matrix  for the signal and idler modes can be expressed as

\begin{widetext}
	\begin{align}
		\rho_{s,i}\left(\tau\right) = \sum_{\{N_{s},N_{i}\}}\sum_{\{N'_{s},N'_{i}\}}&|C^{\left(0\right)}_{N_{s},N_{i}} C^{\left(0\right)}_{N'_{s},N'_{i}}|e^{i\left(N_{s}-N'_{s}\right)\theta_{s}+i\left(N_{i}-N'_{i}\right)\theta_{i}}\;\times\nonumber\\
		&\times\Big[1+\tau\Big(2|\alpha_{i}||\alpha_{s}||\gamma|\cos\Phi-\frac{|\gamma|}{|\alpha_{s}||\alpha_{i}|}\left(N_{s}N_{i}e^{i\Phi}+N'_{s}N'_{i}e^{-i\Phi}\right)\Big)\Big]\ket{N_{s},N_{i}}\bra{N'_{s},N'_{i}},
		\label{eqn:density}
	\end{align}
\end{widetext} 
 
\begin{figure}
	\includegraphics[width=\linewidth,keepaspectratio]{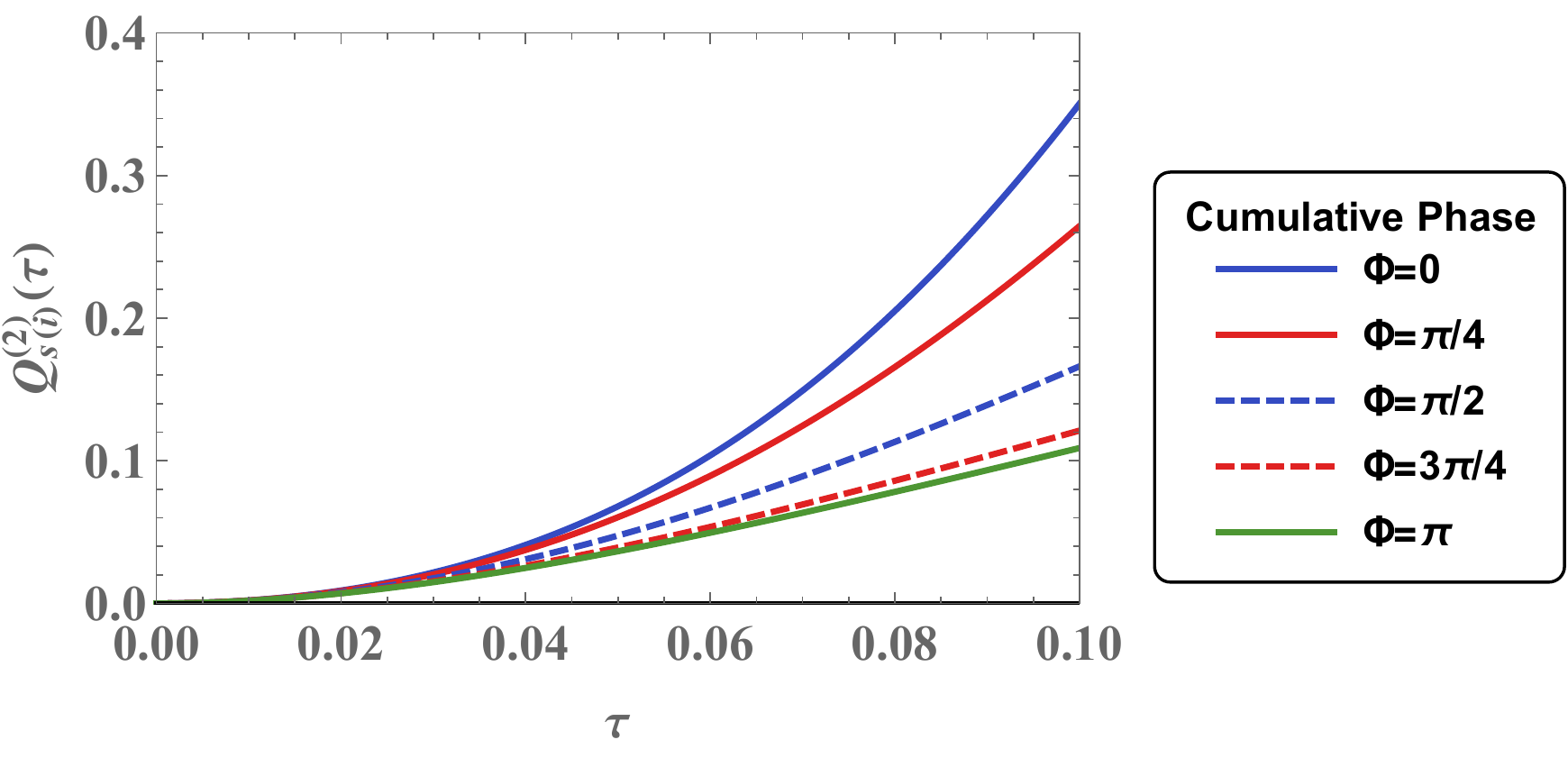}
	\caption{The Mandel $Q$ parameter calculated to second order in $\tau$ for the signal(idler) mode, with $|\alpha_{s}|=|\alpha_{i}|=\sqrt{3}\;\text{and}\;|\gamma|=\sqrt{10}$. For short times, the pump remains Poissonian while the signal/idler modes become super-Possonian.  The effects of the cumulative phase on the short term evolution of the signal/idler mode photon statistics are shown.}
	\label{fig:MandelPTs}
\end{figure}
\begin{figure}
	\centering
	\hspace*{-0.2cm}
	\subfloat[][]{\includegraphics[width=1.0\linewidth,keepaspectratio]{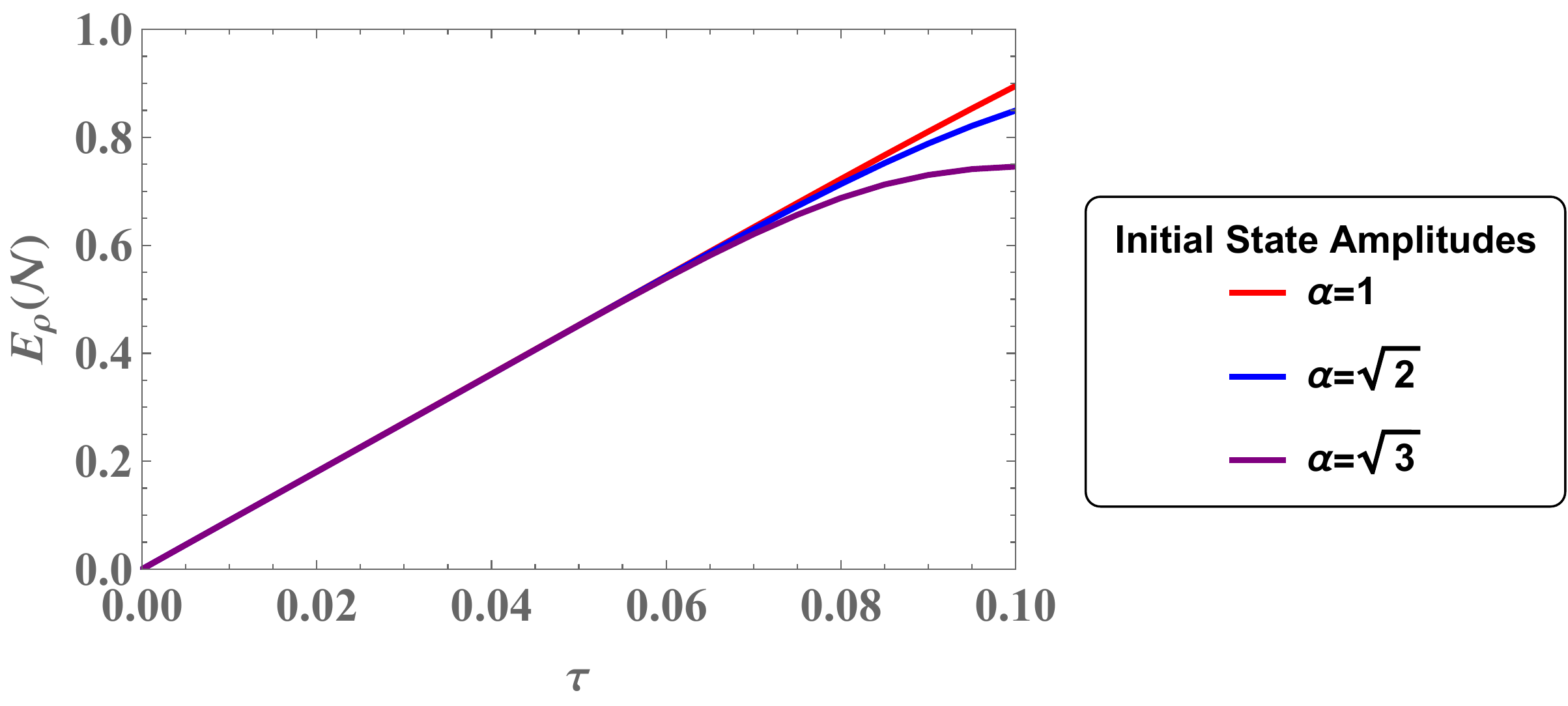}
		\label{fig:PT_LogNegAmp}}
	\\
	\hspace*{-0.83cm}
	\subfloat[][]{\includegraphics[width=0.92\linewidth,keepaspectratio]{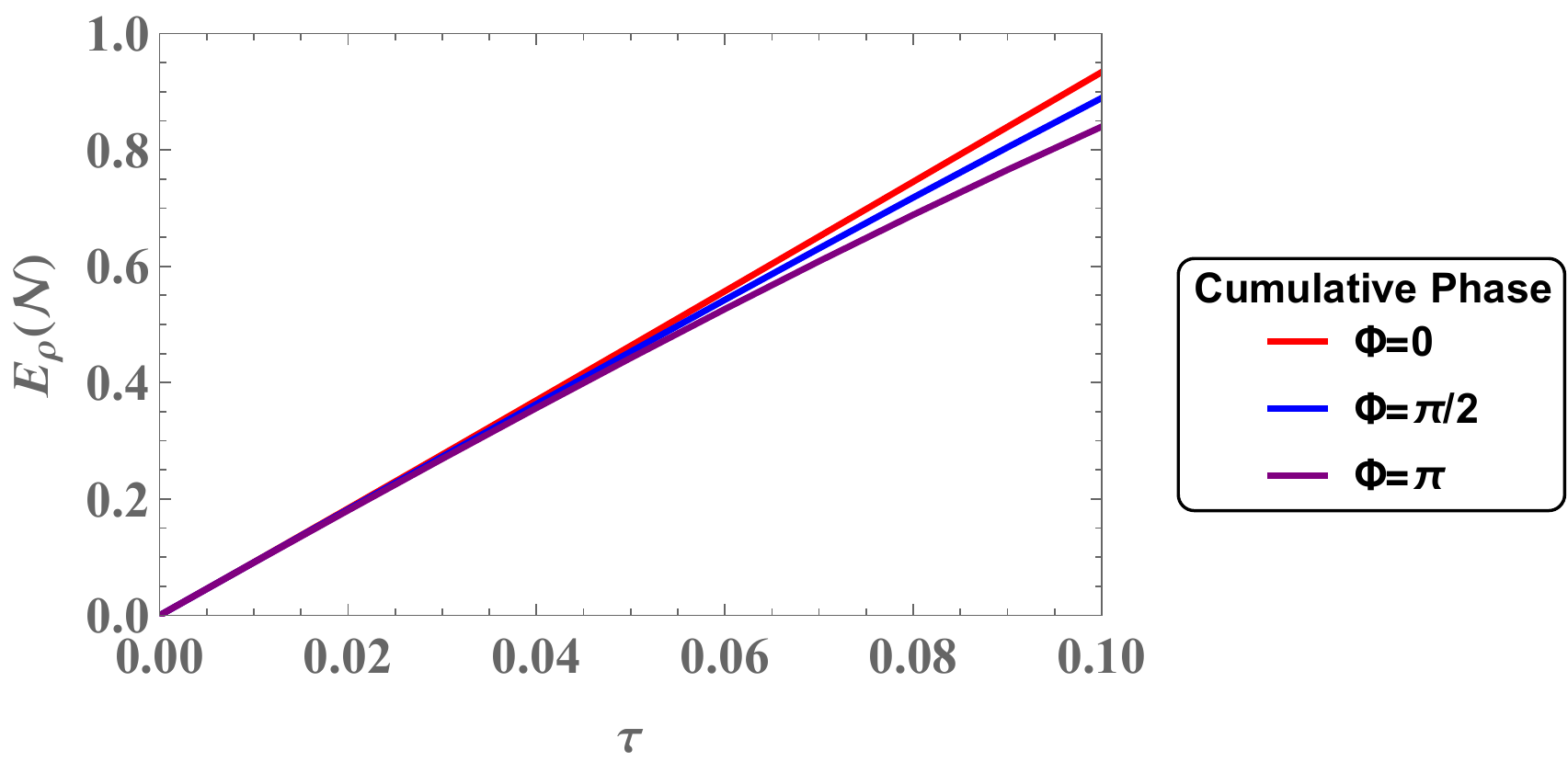}
		\label{fig:PT_LogNegPhase}}
	\caption{\ref{fig:PT_LogNegAmp}.) Log negativity versus scaled time $\tau$ for $\Phi=0$ with $|\alpha_{s}|=|\alpha_{i}|=\alpha$ and $|\gamma|=\sqrt{10}$ and \ref{fig:PT_LogNegPhase} the log negativity for various cumulative phase $\Phi$ with $|\alpha_{s}|=|\alpha_{i}|=\sqrt{3}$ and $|\gamma|=\sqrt{10}$.  Note that the entanglement between signal and idler modes will, in general, depend on the amplitudes of the seeded coherent states as well as the value of the cumulative phase $\Phi$.  This is not true for the case of a constant pump.}
	\label{fig:PT_LogNeg} 
	\end{figure}

 
 
\noindent where $C^{\left(0\right)}_{n_{s},n_{i}}$ is the product of the initial signal and idler mode coherent state coefficients, given by Eq. \ref{eqn:10}.  In general, the matrix elements will depend on the individual phases, as the initial state will.  However, corrections to higher order in time, as well as quantities dependent solely on the diagonal elements of the reduced density matrix (such as the Mandel $Q$ parameter discussed above), will depend only on the cumulative phase $\Phi$. Diagonalization of the two-mode reduced density matrix for the signal and idler modes yields elements $\rho_{kk}$ that are solely dependent on the cumulative phase $\Phi$ and not the individual phases. Consequently, as far as phase dependency is concerned, the logarithmic negativity is strictly a function of $\Phi$.  It is also important to note that higher order corrections will also depend on the initial state amplitudes.  This is demonstrated in Fig. \ref{fig:PT_LogNegAmp} where we plot the logarithmic negativity, Eq. \ref{eqn:log1}, to third order in $\tau$ for different values of initial signal/idler coherent state amplitude with a constant phase. We also plot the logarithmic negativity for different values of $\Phi$ with constant coherent state amplitudes in Fig. \ref{fig:PT_LogNegPhase}. While all three modes remain nearly coherent states, the logarithmic negativity increases linearly with time and does not show any dependency on the initial coherent state  amplitudes nor on the choice of cumulative phase.  This is precisely what one would expect for the case of a constant pump where $r=|\gamma|t\sim\sqrt{n_{p}}\tau$. That is, the logarithmic negativity is linear in time $\tau$ for a constant pump amplitude. This is sensible, as the parametric approximation still holds for very short times. For later times, as the pump occuption varies, this no longer remains true, and we begin to see dependency on other state parameters. We note that a similar expression to Eq. (\ref{eqn:logneg1}) can be obtained for the reduced density matrix of the signal and pump modes by tracing out the idler mode.  

It is worth noting here that the logarithmic negativity is a measure of bipartite entanglement \cite{ref:PPT2}. In using the logarithmic negativity to measure entanglement between the signal and idler modes in this case, we must trace out the pump mode, which in turn throws out information about the total state as the pump can be shown to be entangled with the signal/idler modes for times $\tau>0$. We will consider a means of working around this in the following section by partitioning our tripartite state into two subsystems consisting of the pump and combined signal/idler modes and measuring the total correlations between subsystems.   

\section{\label{sec:level4} IV. Quantizing the pump field - Numerical Analysis}
 
Next, we consider state evolution of the three modes through numerical analysis.  We use a fourth-order Runge Kutta method as our means of numerical integration, where the differential equations to be solved for the state coefficients are determined through the usual time-dependent Schr\"{o}dinger equation,  

\begin{equation}
	\hat{H}_{\text{I}}\ket{\psi\left(t\right)}=i\hbar\frac{\partial}{\partial t}\ket{\psi\left(t\right)},
	\label{eqn:33}
\end{equation}

\noindent where the time-dependent state has the form

\begin{equation}
\ket{\psi\left(t\right)} = \sum_{n_{p},n_{s},n_{i}}^{\infty} C_{n_{p},n_{s},n_{i}}\left(t\right)\ket{n_{p}}_{p}\ket{n_{s}}_{s}\ket{n_{i}}_{i},
\label{eqn:34}
\end{equation}

\noindent and where the state vectors $\ket{n_{p}}_{p}\ket{n_{s}}_{s}\ket{n_{i}}_{i}$ denote the pump, signal and idler modes, respectively.  Plugging Eq.(\ref{eqn:34}) into Eq.(\ref{eqn:33}) yields the differential equation

\begin{align}
	\dot{C}_{n_{p,n_{s},n_{i}}}=\kappa&\sqrt{n_{p}\left(n_{s}+1\right)\left(n_{i}+1\right)}\;C_{n_{p}-1,n_{s}+1,n_{i}+1} - \nonumber\\
	& -\kappa\sqrt{\left(n_{p}+1\right)n_{s}n_{i}}\;C_{n_{p}+1,n_{s}-1,n_{i}-1},
	\label{eqn:35}
\end{align}  

\noindent subject to the intial value condition

\begin{figure}
	\centering
	\hspace*{-0.6cm}
	\subfloat[][]{\includegraphics[width=1.05\linewidth,keepaspectratio]{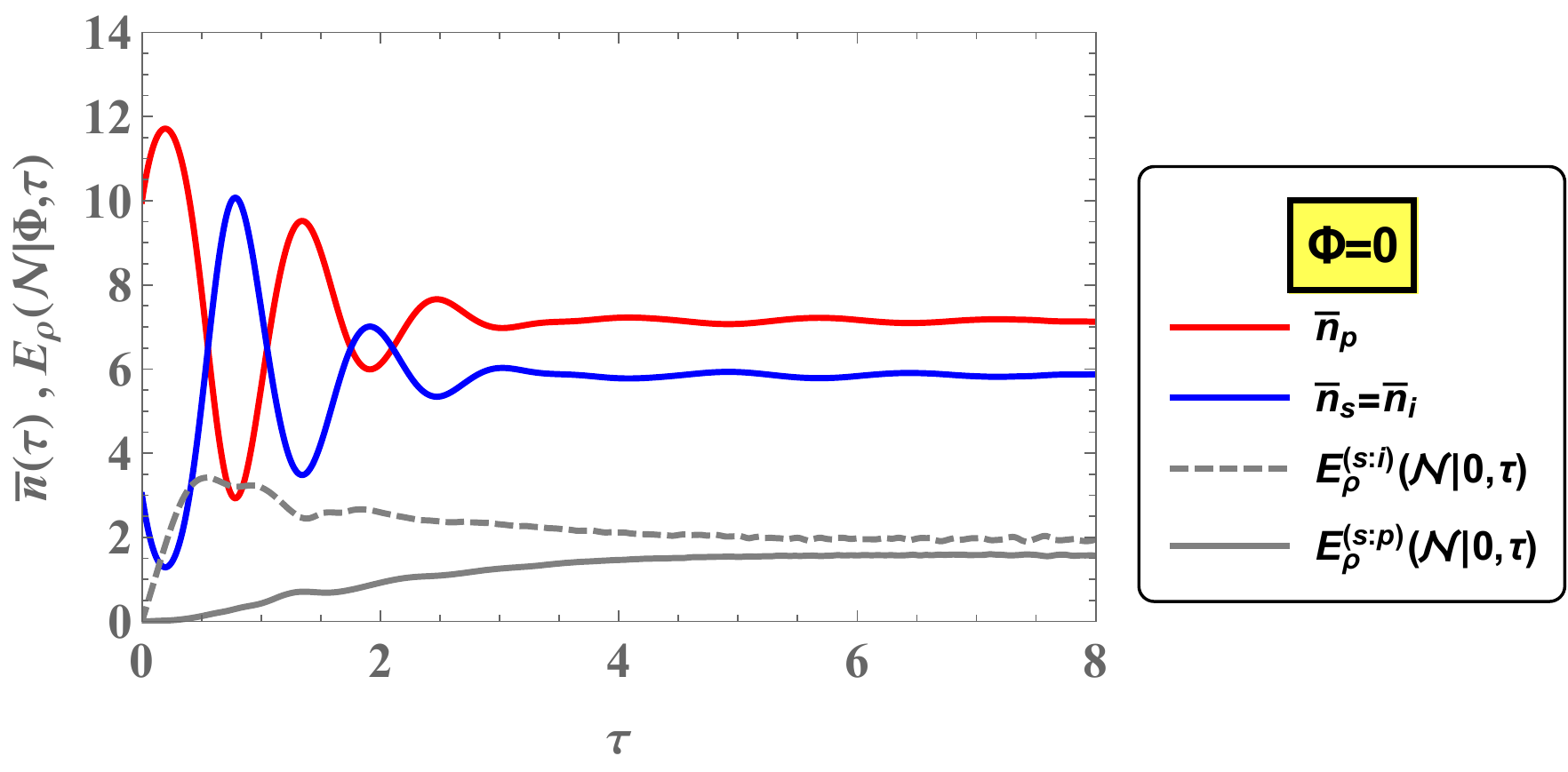}
		\label{fig:NUM_AvgPhoton0}}
	\\
	\hspace*{-0.6cm}
	\subfloat[][]{\includegraphics[width=1.07\linewidth,keepaspectratio]{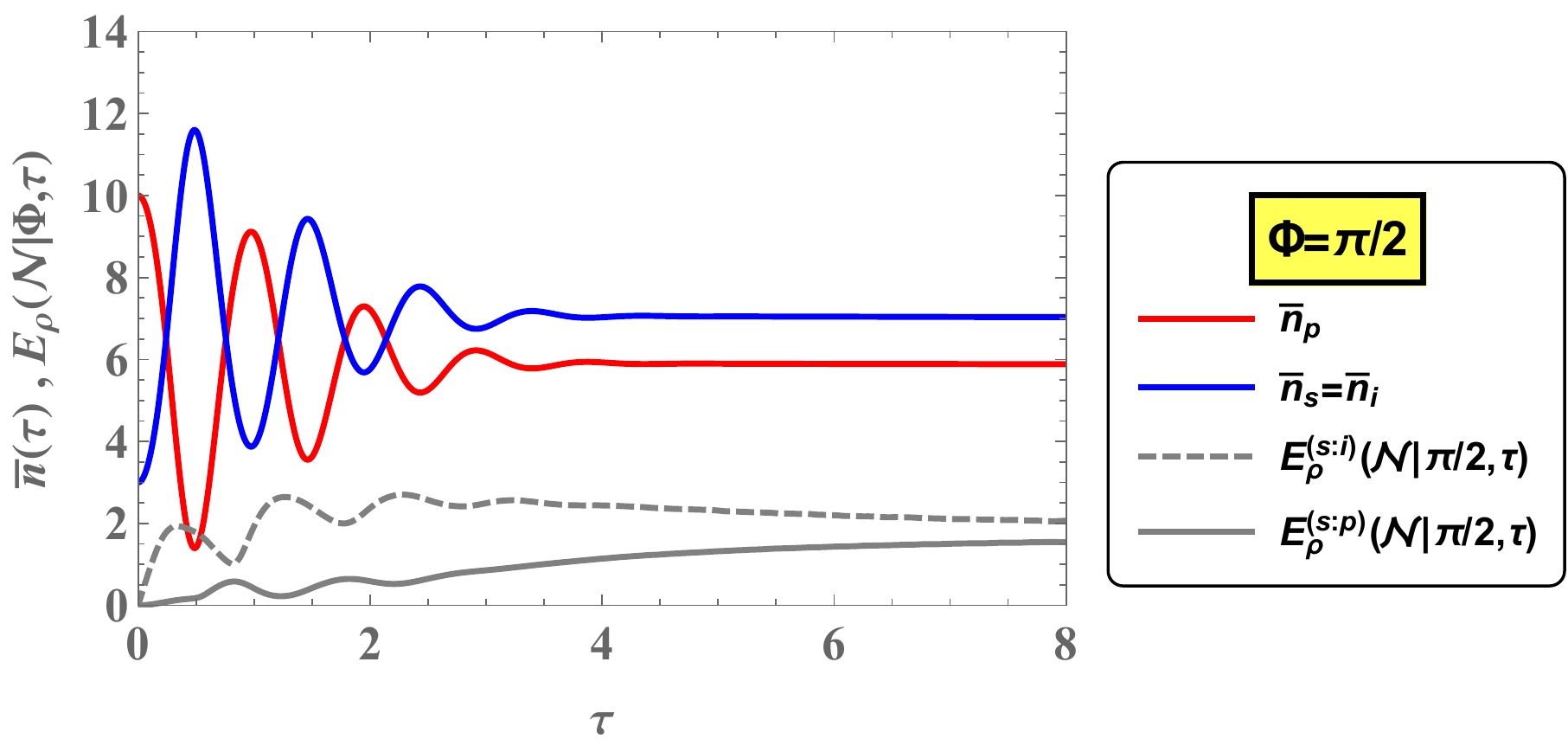}
		\label{fig:NUM_AvgPhotonPiOver2}}
	\\
	\hspace*{-0.75cm}
	\subfloat[][]{\includegraphics[width=1.025\linewidth,keepaspectratio]{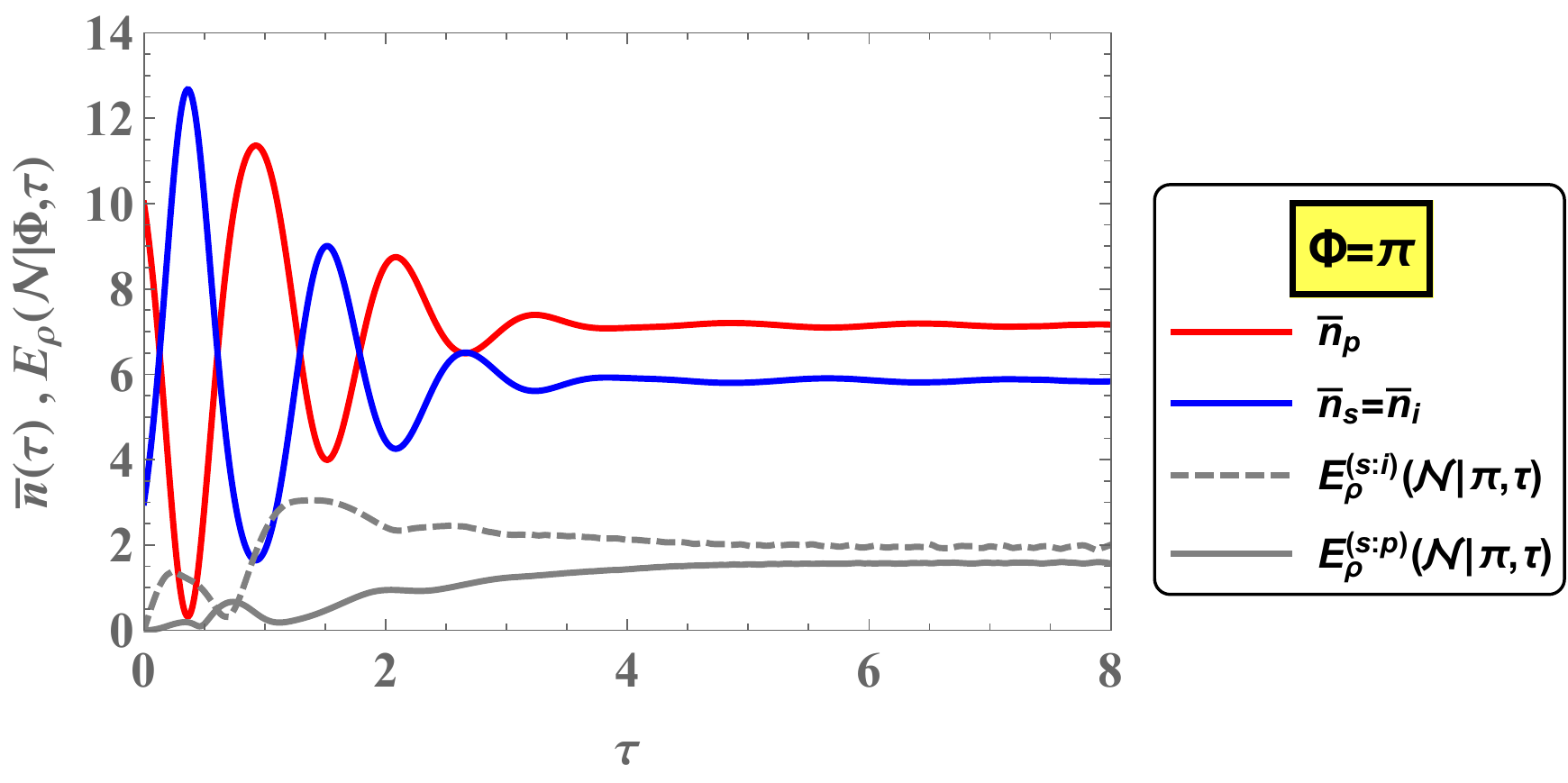}
		\label{fig:NUM_AvgPhotonPi}}
	\caption{Average Photon Numbers for the pump (red) and signal/idler (blue) modes for \ref{fig:NUM_AvgPhoton0}.) $\Phi=0$, \ref{fig:NUM_AvgPhotonPiOver2}.) $\Phi=\pi/2$ and \ref{fig:NUM_AvgPhotonPi}.) $\Phi=\pi$.  We include the logarithmic negativity for one set of individual phase values between signal/idler modes and signal/pump modes for reference.}
	\label{fig:NUM_AvgPhoton} 
\end{figure}

\begin{align}
	C_{n_{p},n_{s},n_{i}}\left(0\right)=&e^{-\tfrac{1}{2}\left(|\alpha|^{2} + |\beta|^{2} + |\gamma|^{2}\right)}\times\frac{|\alpha|^{n_{s}}|\beta|^{n_{i}}|\gamma|^{n_{p}}}{\sqrt{n_{s}!n_{i}!n_{p}!}}\times \nonumber \\
	& \times e^{i\left(n_{s}\theta_{1}+n_{i}\theta_{2}+2n_{p}\phi\right)},
	\label{eqn:36}
\end{align}

\noindent which are simply a product of the initial signal/idler/pump coherent state coefficients prescribed by Eq. (\ref{eqn:10}).  From Eq. (\ref{eqn:35}), the state coefficients and corresponding density matrix can be determined and used to characterize the evolution of the state statistics. 

\begin{figure}
	\centering
		\hspace*{-0.25cm}
	\subfloat[][]{\includegraphics[width=1.0\linewidth,keepaspectratio]{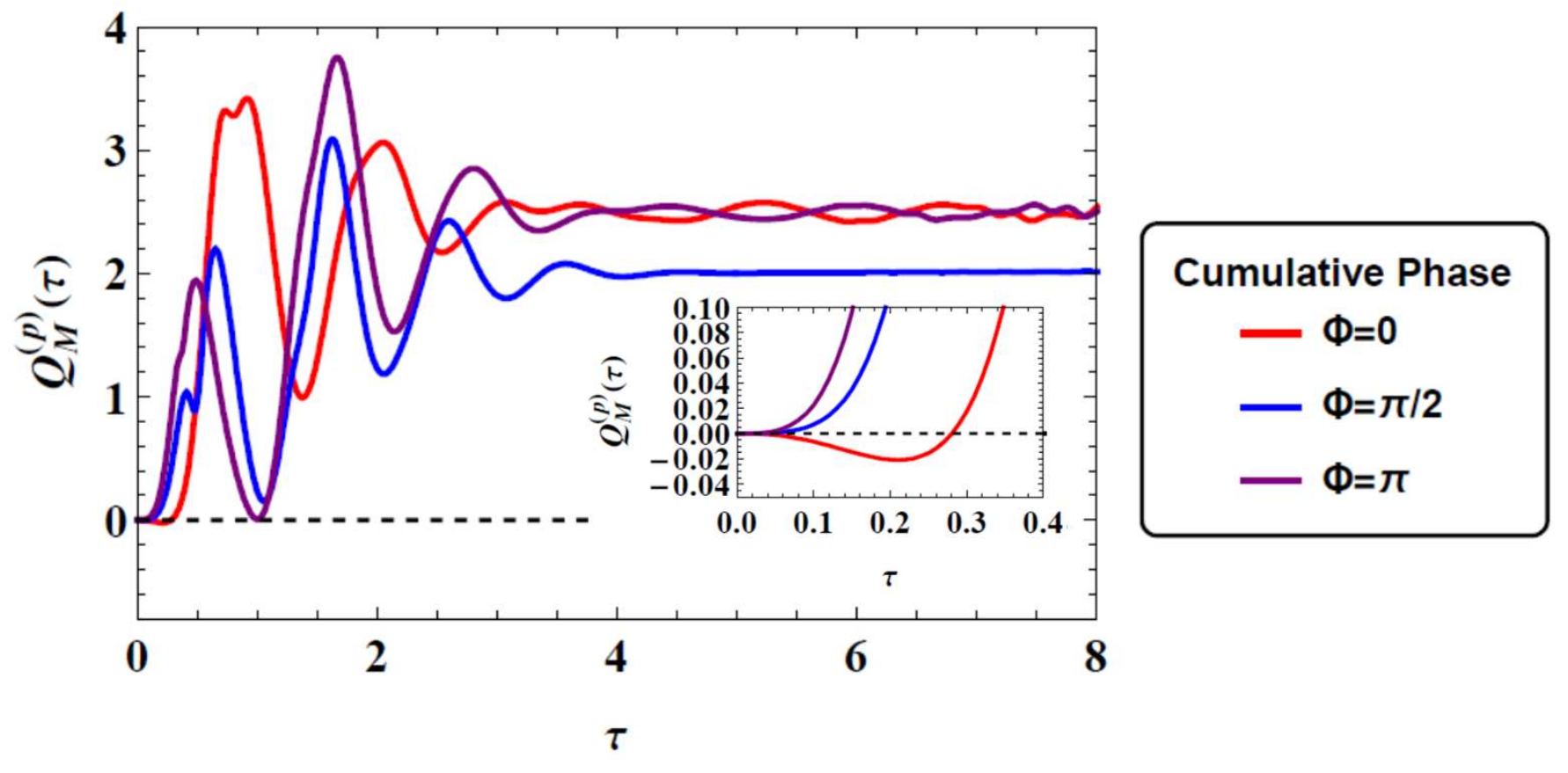}
		\label{fig:Num_Qp}}
	\\
	\hspace*{-0.4cm}
	\subfloat[][]{\includegraphics[width=1.02\linewidth,keepaspectratio]{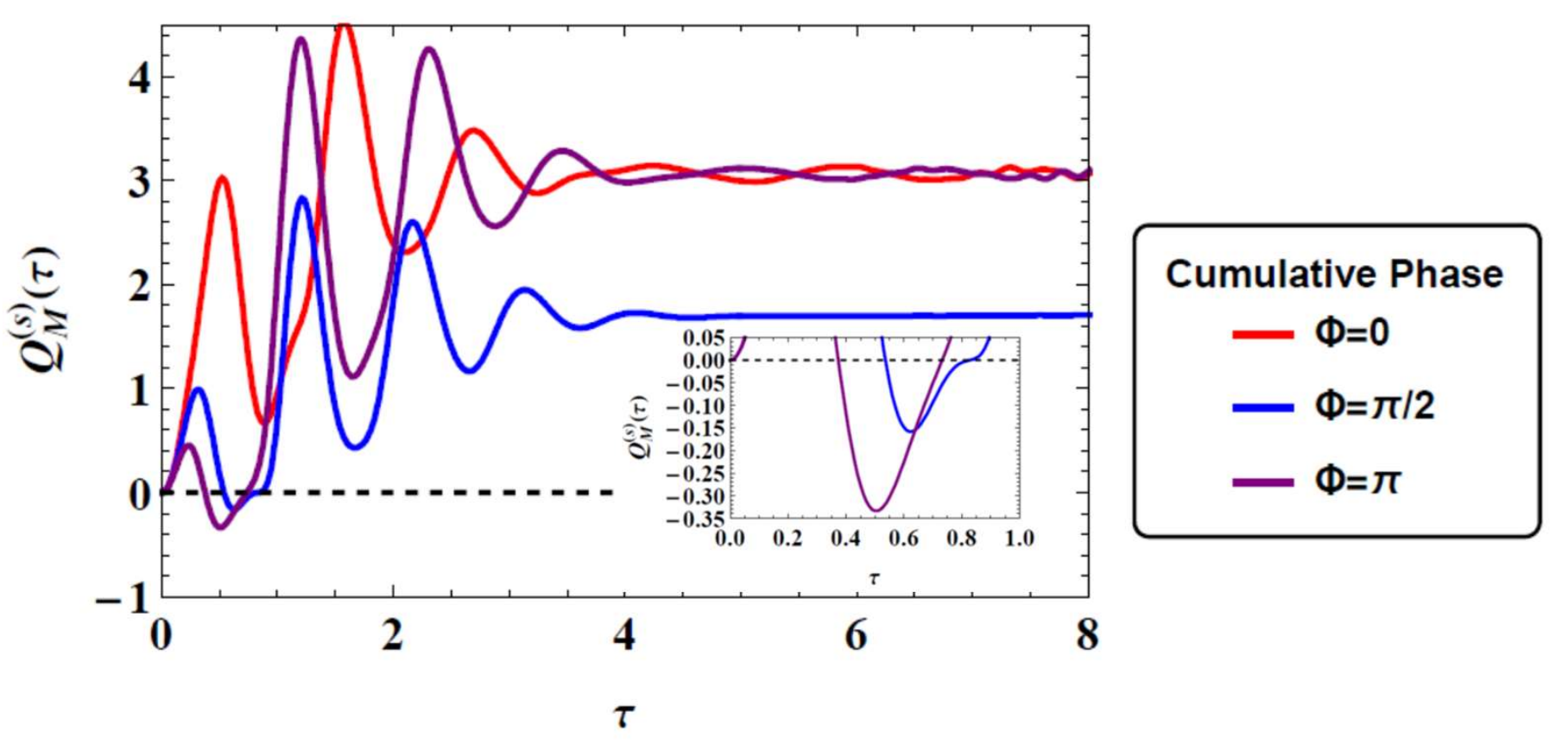}
		\label{fig:Num_Qs}}
	\caption{Mandel $Q$ parameter for \ref{fig:Num_Qp}.) the pump mode and \ref{fig:Num_Qs}.) the signal mode, for several choices of the phase $\Phi$. During transient times, both modes can display sub-Poissonian statistics.}
	\label{fig:Num_MandQ} 
\end{figure}

We begin with a discussion of the state statistics. In Fig. (\ref{fig:NUM_AvgPhoton}) we plot the pump, signal/idler average photon numbers for a selection of values of the cumulative phase $\Phi$, with the corresponding signal-idler logarithmic negativity. During transient times, back action within the cavity causes the average photon number for each mode to oscillate, with the greatest joint-signal/idler occupation average, corresponding to the largest conversion rate (and for our case, maximal pump depletion), occurring for $\Phi=\pi$, consistent with what one would see for the case of a constant pump.  Interestingly, we see that points in which the average occupation between the pump and signal(idler) modes cross closely correspond to points of local extrema in the signal-idler logarithmic negativity.  In Fig. \ref{fig:Num_MandQ} we plot the corresponding pump/signal(idler) $Q$ parameter for $\Phi=0,\pi/2,\pi$. For the case of $\Phi=0$, the pump quickly becomes sub-Poisonnian as the initial back action from the signal/idler modes result in an increase in average photon number, before becoming super-Poissonian at longer times.  Meanwhile, the signal(idler) also become sub-Poissonian, with $Q$ becoming minimum at a point near the second crossing of the average mode occupations.  The $Q$ for all three modes asymptote at long times, suggesting the steady state has super-Poissonian statistics (discussed in more depth in Section V), however, the steady state statistics \textit{will} depend on the value of the cumulative phase $\Phi$. 

\begin{figure}
	\centering
	\hspace*{-0.5cm}
	\subfloat[][]{\includegraphics[width=1.07\linewidth,keepaspectratio]{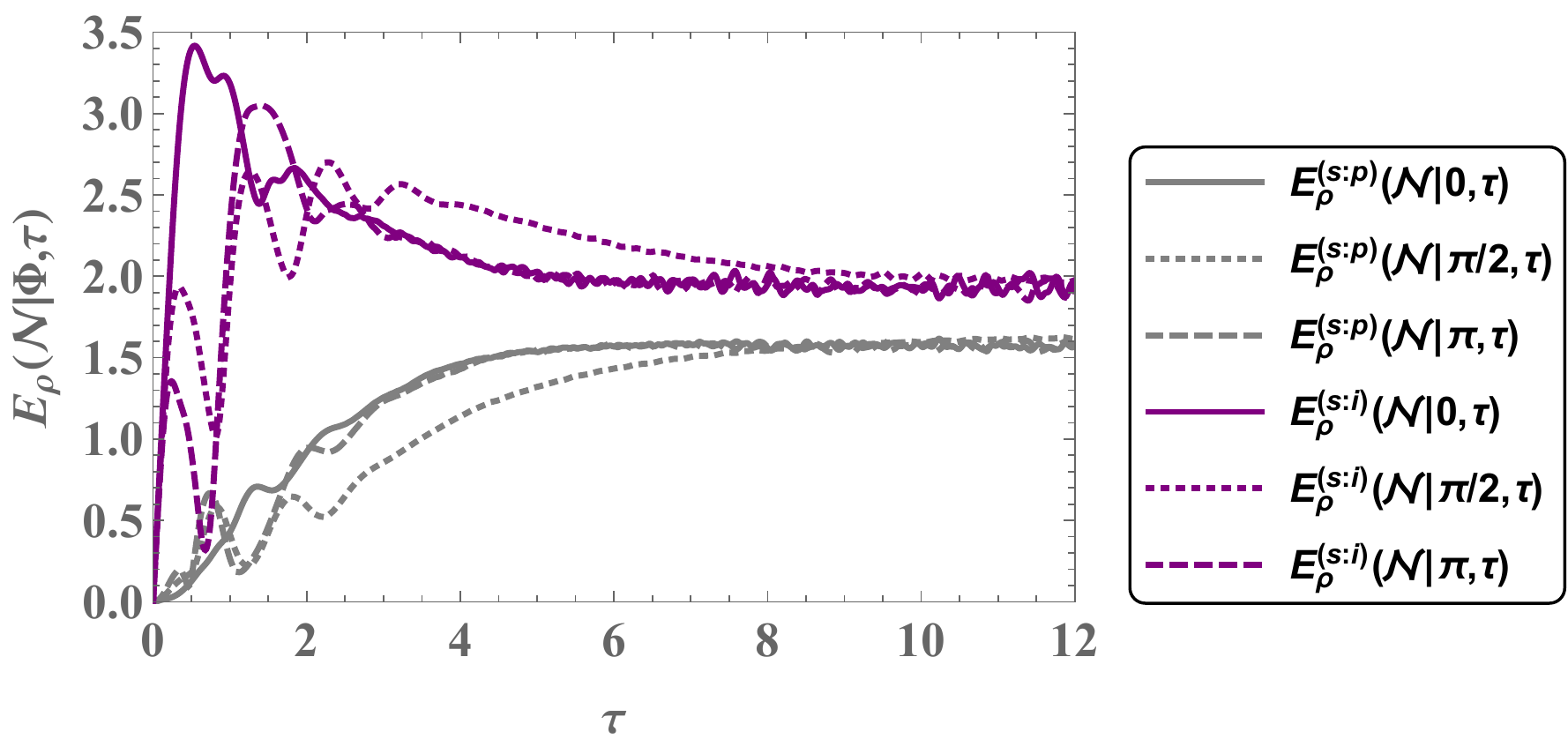}
		\label{fig:Num_LogNeg}}
	\\
	\hspace*{-0.2cm}
	\subfloat[][]{\includegraphics[width=1.0\linewidth,keepaspectratio]{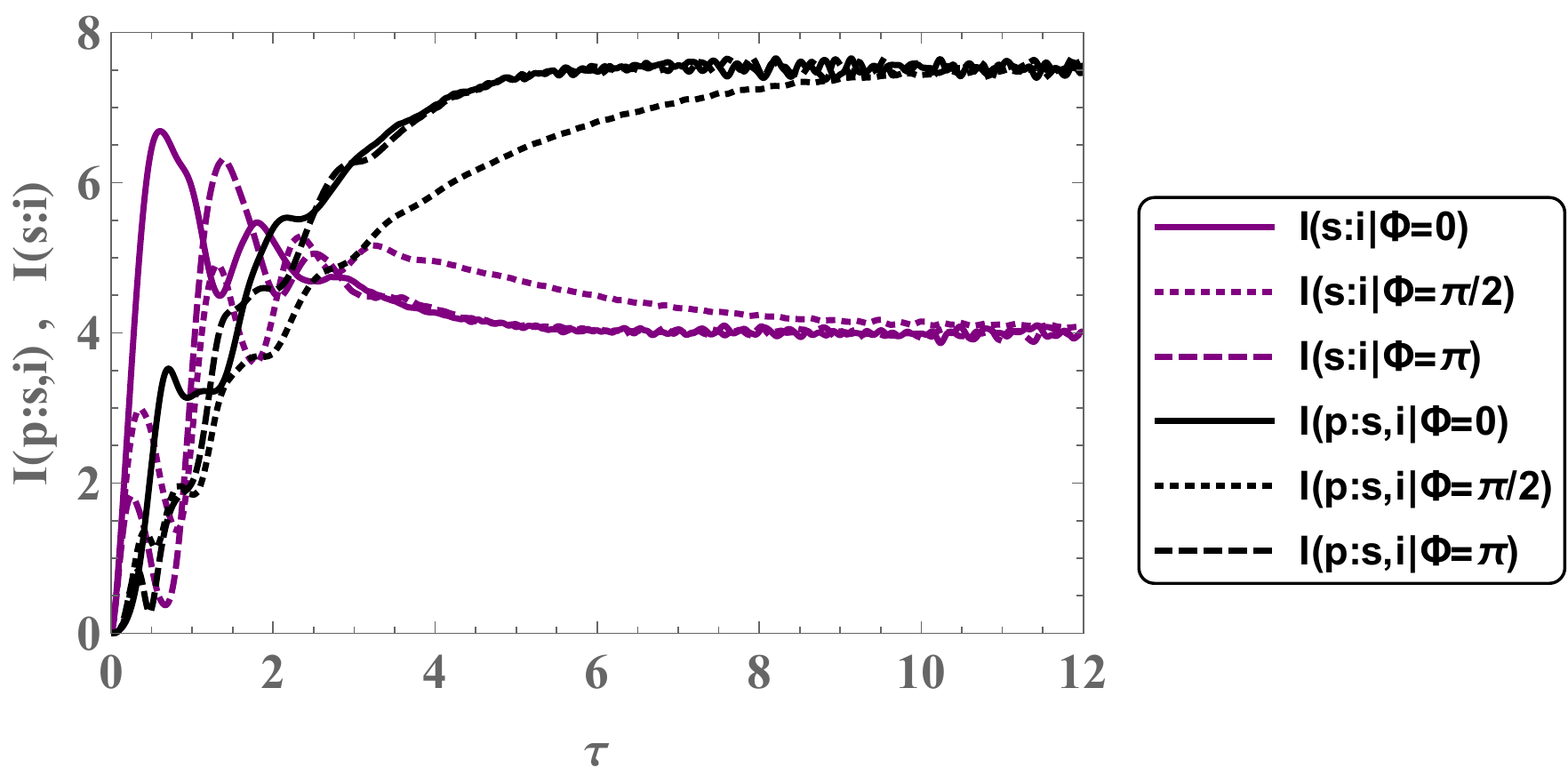}
	\label{fig:Num_MutInfo}}
	\caption{\ref{fig:Num_LogNeg}.) Log Negativity $E_{\rho}\left(\mathcal{N}|\Phi,\tau\right)$ between both signal and idler modes as well as signal and pump modes. \ref{fig:Num_MutInfo}.) Mutual information between signal and idler subsystems and between pump and signal/idler subsystems. }
	\label{fig:Num_Entang} 
\end{figure}

As we have stated earlier, one can investigate entanglement in the three mode state using bipartite measures by partitioning the state into two separate subsystems: one consisting of the pump mode and the other encompassing both signal and idler modes.  This bipartite split allows us to measure the total correlations, both classical and quantum, between subsystems \cite{ref:Mutualref1} by calculating the mutual information \cite{ref:Mutualref2}, given by 

\begin{equation}
	I\left(p:s,i\right)= S_{p}+S_{s,i}-S_{p,s,i},
	\label{eqn:Mutual1}
\end{equation}    

\noindent where $S_{i}$ denotes the entropy of the $i^{\text{th}}$ (sub)system. Since the state remains pure at all times, $S_{p,s,i} \equiv 0$ and the subsystem entropies satisfy $S_{p}=S_{s,i}$.  Consequently, the entropy of the pump mode is a direct measure of the entanglement between pump and signal/idler subsystems, $I\left(p:s,i\right)=2S_{p}$.  Likewise, we can use the mutual information to comment on the entanglement between signal and idler modes 

\begin{equation}
	I\left(s:i\right)=S_{s}+S_{i}-S_{s,i}=S_{s}+S_{i}-S_{p},
	\label{eqn:Mutual2}
\end{equation} 

\noindent where we have used the relation between subsystem entropies detailed prior to Eq. (\ref{eqn:Mutual1}).  We plot the mutual information between pump and signal/idler modes and as well as between signal and idler modes for several values of the phase $\Phi$ in Fig. (\ref{fig:Num_MutInfo}).  For transient times, the entanglement for both cases vary greatly depending on the choice of cumulative phase, however, at long times, the mutual informations asymptote to the same value for $\Phi=0,\;\pi$. In general, at long times the entanglement properties between signal and idler modes will depend on the choice of cumulative phase, however entanglement between the signal(idler) modes with the pump is independent of this phase as shown in Fig. \ref{fig:Num_Entang}.  Note that for all values of the phase $\Phi$ at long times, while the signal is more entangled with the idler mode than it is with the pump mode, the total correlations between the pump and the joint signal/idler modes is substantially greater than between the signal and idler modes.  

\section{\label{sec:level5} V. Modeling the Steady State - Long Times}

\begin{figure}
	\centering
	\hspace*{-0.0cm}
	\subfloat[][]{\includegraphics[width=1.03\linewidth,keepaspectratio]{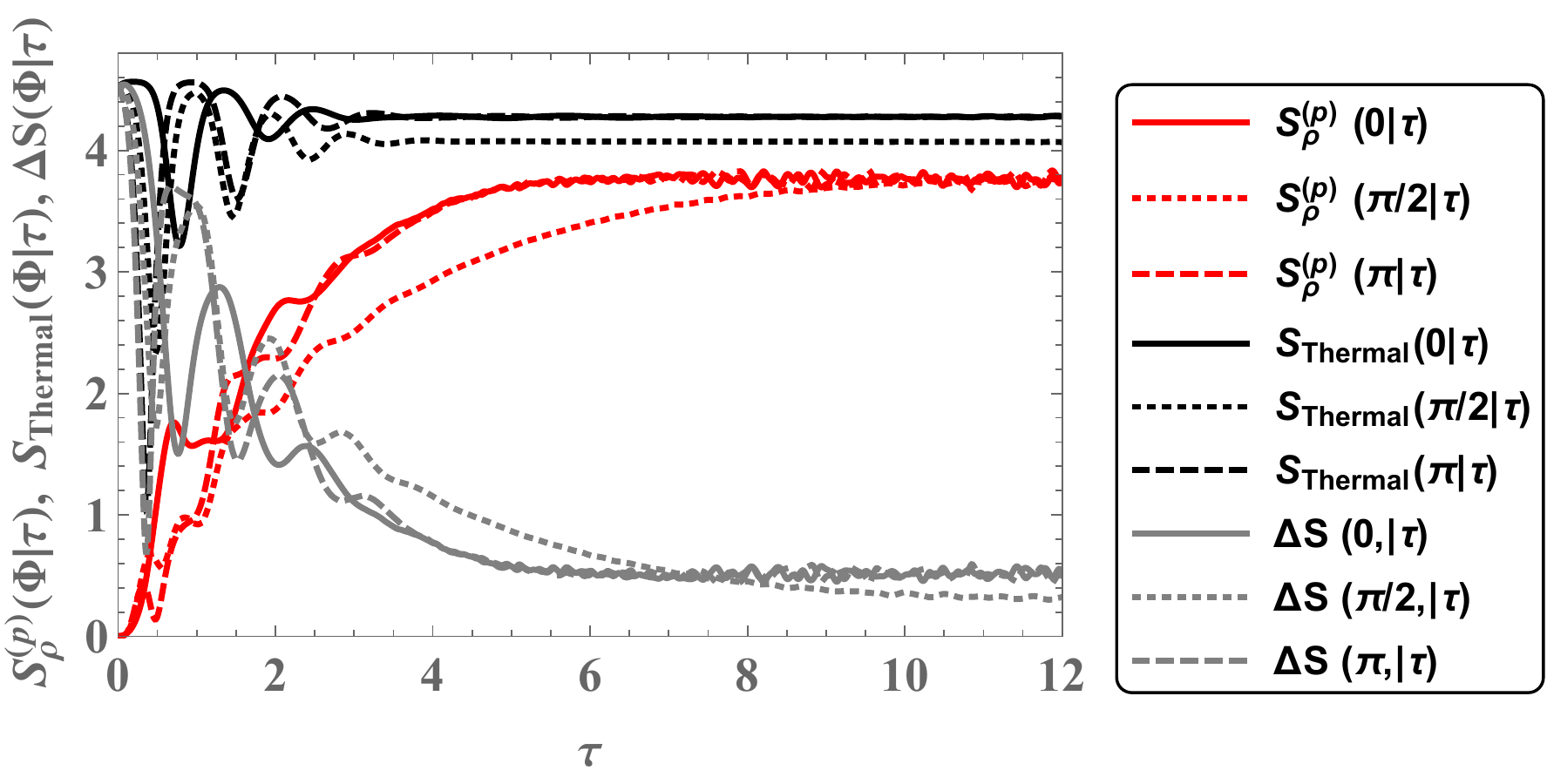}
		\label{fig:Num_EntropyPump}}
	\\
	\hspace*{-0.0cm}
	\subfloat[][]{\includegraphics[width=1.03\linewidth,keepaspectratio]{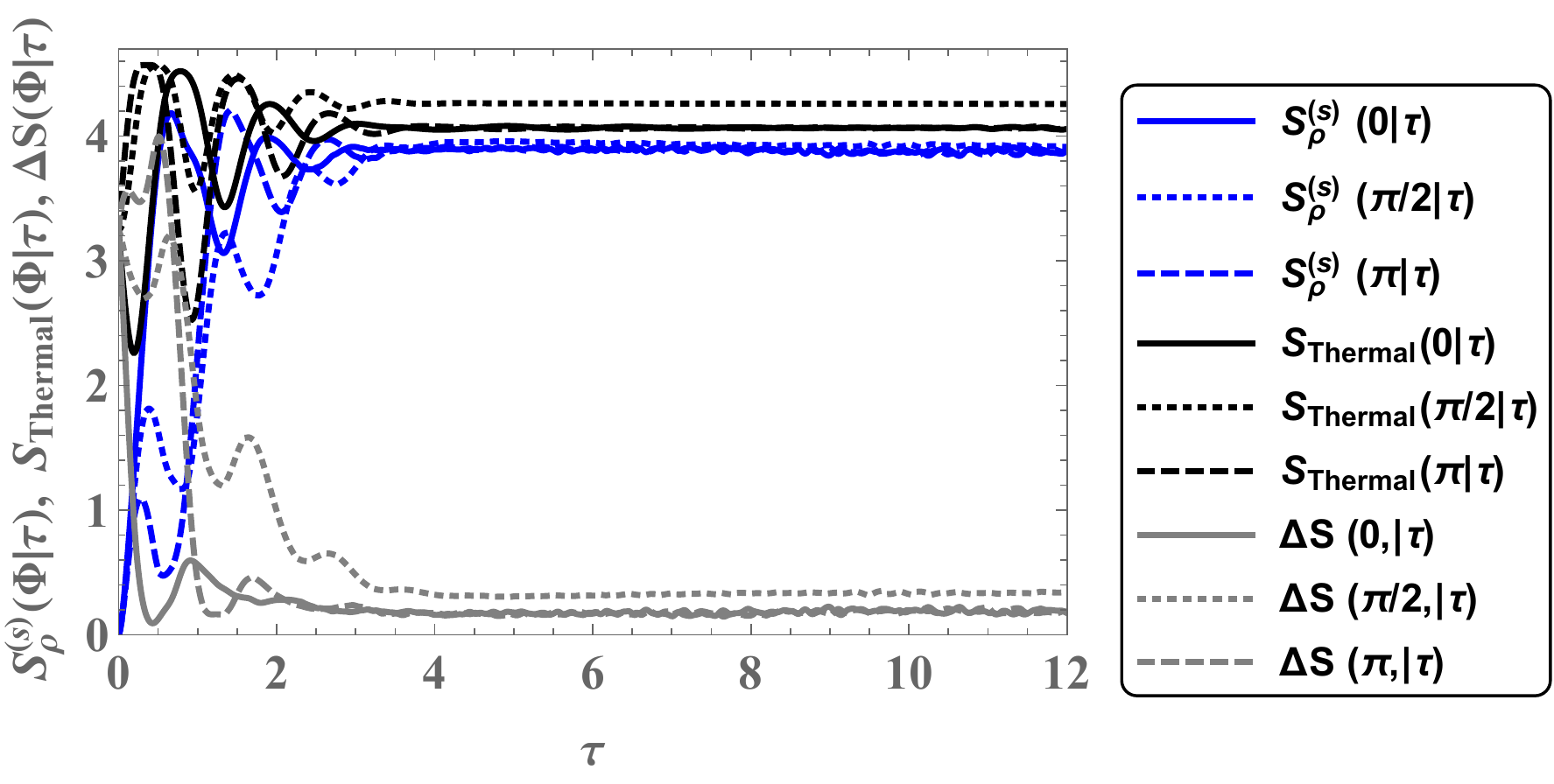}
		\label{fig:Num_EntropySignal}}
	\caption{von Neumann Entropy $S_{\rho}^{\left(j\right)}\left(\Phi|\tau\right)$ for the signal and pump mode, characterized by reduced density $\rho_{s}\left(\tau\right)$ and $\rho_{p}\left(\tau\right)$ respectively, as well as the von Neumann entropy of an effective thermal state of similar occupation number for each case.  For long times, the difference between the entropies, $\Delta S$, closely approaches zero for both modes. This shows that at long times, the signal and pump modes, (but more so the signal/idler modes) are near-thermal. Interestingly, for the choice of $\Phi=\pi/2$, the pump is most thermal-like.}
	\label{fig:Num_EntropyComp} 
\end{figure}

We now turn our attention towards providing a qualitative discussion of the long-time steady state statistics.  We note that the long time dynamics of the trilinear model has been explored in the past for the case of spontaneous parametric down-conversion \cite{ref:LTD1}\cite{ref:LTD2} in which the usual mean-field approximation of the pump mode is made to linearize the boson operator equations.  Our model departs from the previously covered analysis by allowing the pump to remain a quantized mode, thus permitting us to remark on the correlations between the pump and signal/idler modes.  As motivation, we consider a comparison between the von Neumann entropy for the signal/pump modes of the state in question with that of an effective thermal state of equivalent average photon number (at each point in time).  Recall that for a standard two-mode squeezed state, each mode separately has thermal-like statistics \cite{ref:therm1}\cite{ref:therm2}.  This can be verified by, for example, tracing over the idler mode and considering the mixed state occupying the signal mode.  In Fig. \ref{fig:Num_EntropyComp} we plot the von Neumann entropy, defined in Eq. \ref{eqn:neumann}, for both the pump and signal modes, $S_{\rho}^{\left(p\right)}\left(\Phi|\tau\right)$ and $S_{\rho}^{\left(s\right)}\left(\Phi|\tau\right)$, respectively, along with the entropy $S_{\text{Thermal}}\left(\Phi|\tau\right)$ of an effective thermal state of equivalent average photon number at each point in time. That is, we define the effective thermal state for an arbitrary quantum state $\rho$ as $S_{\text{Thermal}}\left(\rho\right)\equiv \sum_{n=0}^{\infty}\left[\bar{n}/\left(\bar{n}+1\right)\right]^{n}\ket{n}\bra{n}$ where $\bar{n}=\text{Tr}\left[\hat{n}\;\rho\right]$ is the mean number occupation number of $\rho$.  For both figures we include the difference $\Delta S\left(\Phi|\tau\right) = S_{\text{Thermal}}\left(\Phi|\tau\right)-S_{\rho}^{\left(j\right)}\left(\Phi|\tau\right)$ where $j=s,p$.  It is important to note that the effective thermal state entropy $S_{\text{Thermal}}$ we are using as a point of comparison has an implicit dependency on $\Phi$, as the average photon number of each mode of our three mode state depends on the cumulative phase $\Phi$. It is clear from Fig. \ref{fig:Num_Entang} that the signal/idler modes are very nearly thermal at long times. This was first posited in the seventies by Walls and Barakat \cite{ref:WBref}; that is, each mode settles down to nearly-thermal-like at long times.  Furthermore it should be noted that at long times the pump field approaches a constant occuptation number, indicating that our expressions for the two-mode squeezed coherent states with a constant pump, outlined in the earlier sections of this paper and in more detail in \cite{ref:1} should remain a valid means of describing the state statistics, provided the initial state is known (or an effective squeeze parameter determined). This explanation remains incomplete, as well, as the signal/idler modes remain entangled with the pump mode at long times.  Consequently, any two-mode analysis will, by nature, be ignoring information lost by tracing over the pump mode.  For the long time regime, the state coefficients are determined through the three-dimensional recursion relation

\begin{align}
	\sqrt{n_{p}\left(n_{s}+1\right)\left(n_{i}+1\right)}&\;C_{n_{p}-1,n_{s}+1,n_{i}+1}= \nonumber \\	
	&\sqrt{\left(n_{p}+1\right)n_{s}n_{i}}\;C_{n_{p}+1,n_{s}-1,n_{i}-1}.
	\label{eqn:ss1}
\end{align}         

\noindent This is not readily soluble.   Making a somewhat crude approximation of $n_{p}\approx n_{p}\pm 1$ allows us to factor out the pump mode in order to qualitatively discuss the signal/idler modes. We point out once again that since the pump mode is entangled with the signal/idler modes, this approximation will effectively be an incomplete treatment of the resulting two-mode state; however, since our goal is to qualify the joint signal/idler two-mode state, it remains a valid means of analysis. We note here that the steady state statistics of the signal(idler) modes, particularly the entanglement properties, \textit{will} depend on the choice of the cumulative phase $\Phi$.  Now the recursion relation simplifies down to

\begin{equation}
	\sqrt{\left(n_{s}+1\right)\left(n_{i}+1\right)}\;C_{n_{s}+1,n_{i}+1}=\sqrt{n_{s}n_{i}}\;C_{n_{s}-1,n_{i}-1}.
	\label{eqn:ss2}
\end{equation}

\begin{figure*}
	\centering
	\hspace*{-0.5cm}
	\subfloat[][]{\includegraphics[width=0.450\linewidth,keepaspectratio]{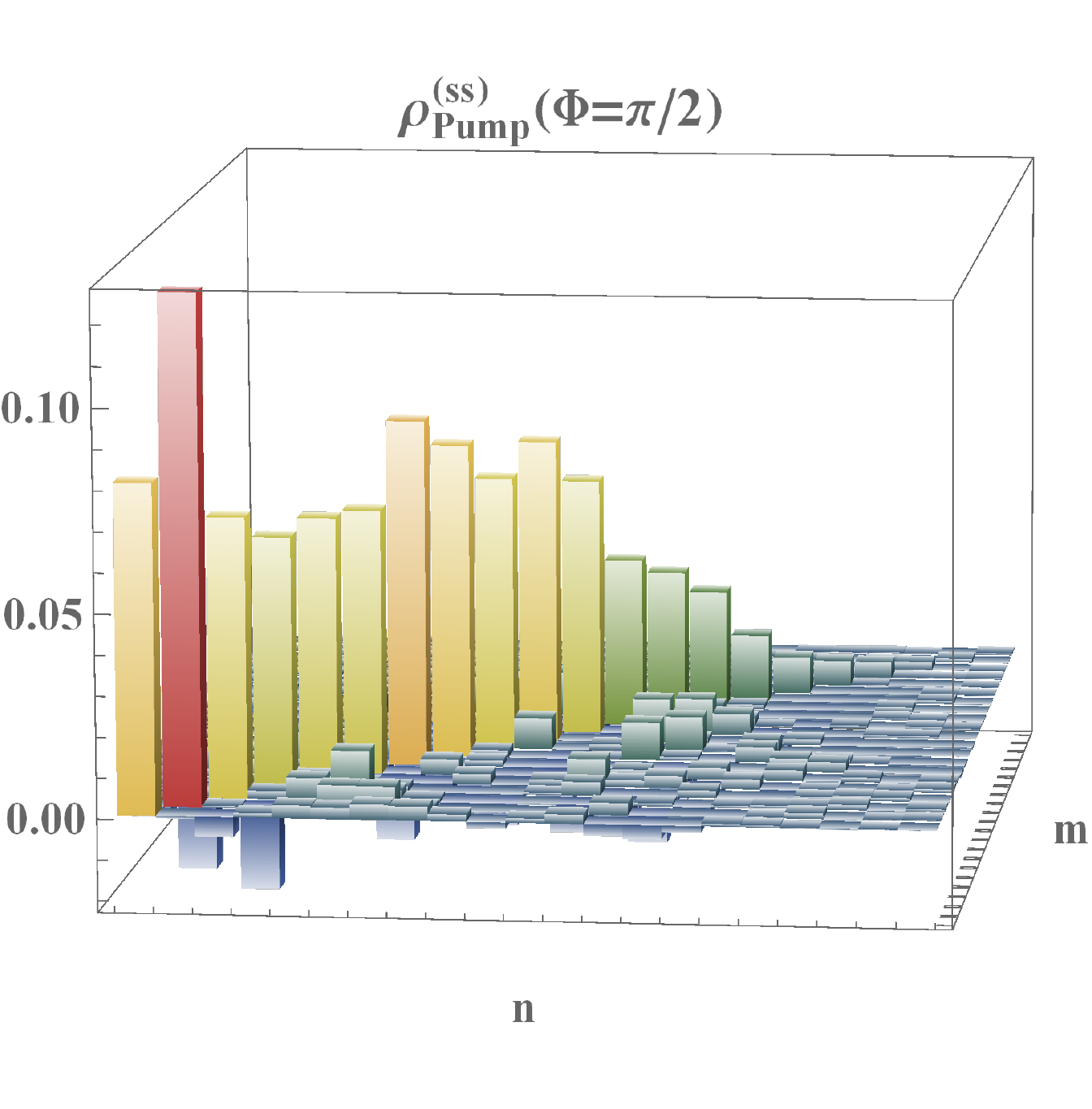}
	\label{fig:Num_DensityPumpPhiPiOver2}}
	\hspace*{1.5cm}
	\subfloat[][]{\includegraphics[width=0.45\linewidth,keepaspectratio]{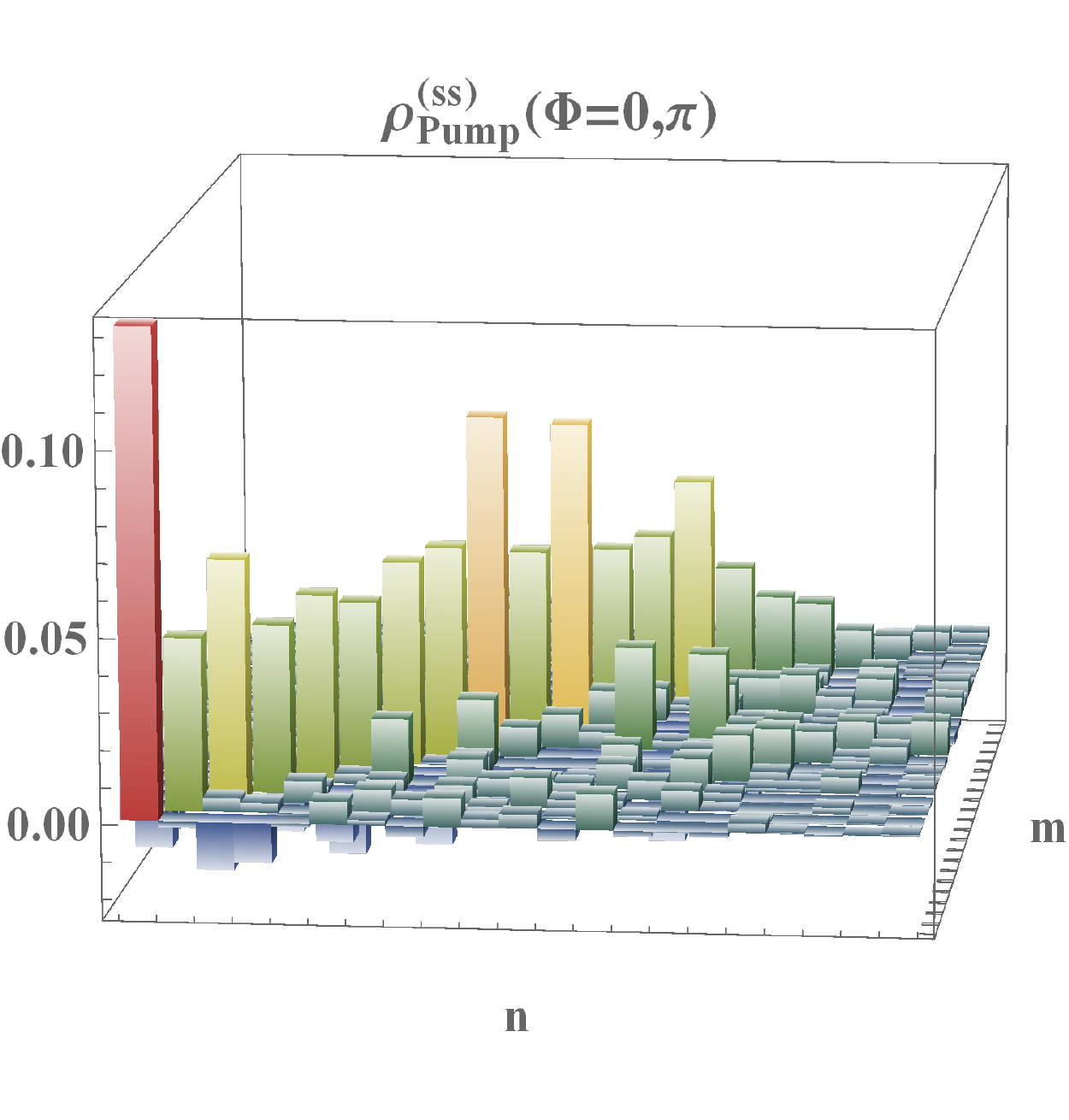}
	\label{fig:Num_DensityPumpPhiPi}}
	\\
	\hspace*{-0.5cm}
	\subfloat[][]{\includegraphics[width=0.450\linewidth,keepaspectratio]{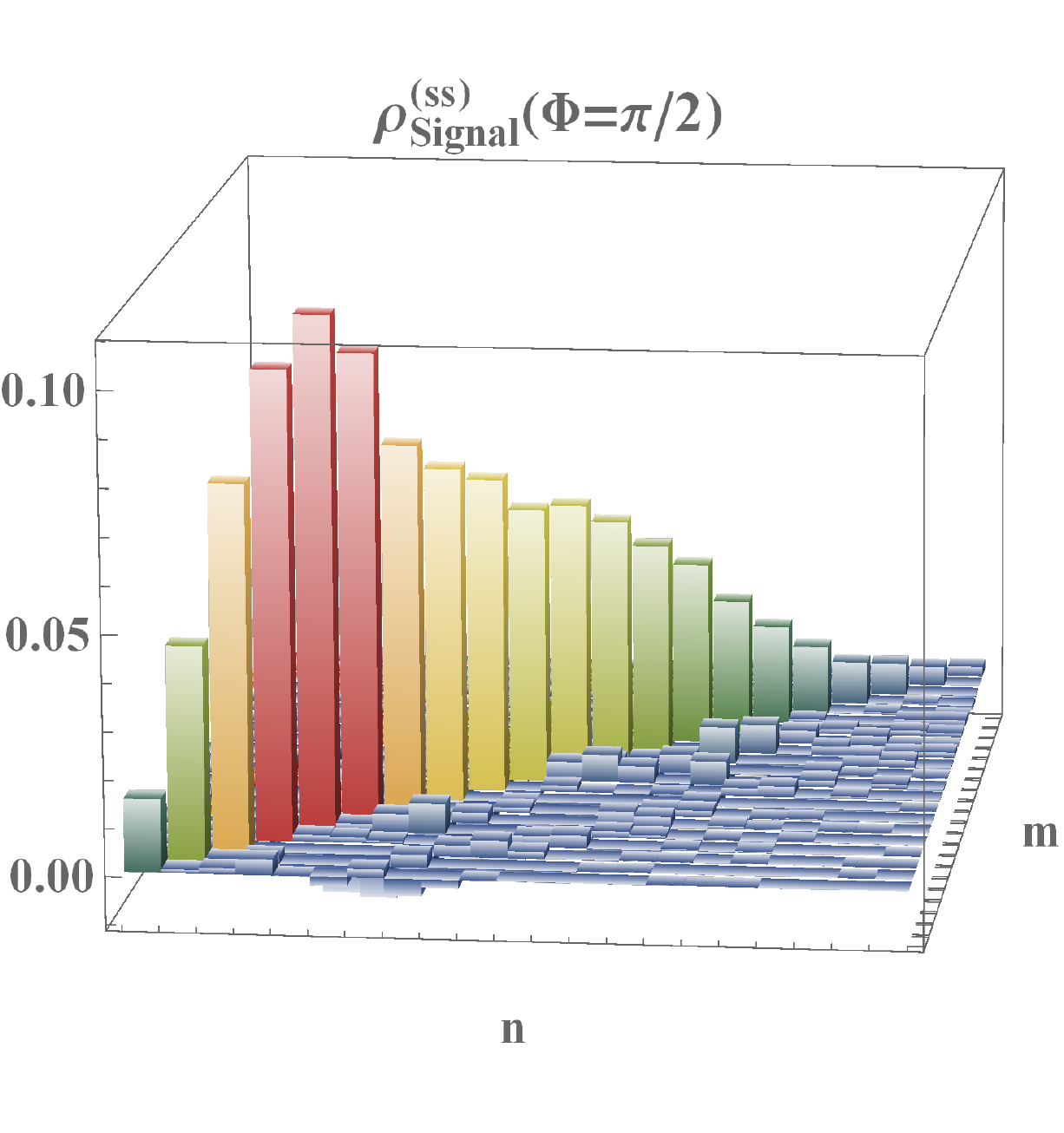}
	\label{fig:Num_DensitySignalPhiPiOver2}}
	\hspace*{1.5cm}
	\subfloat[][]{\includegraphics[width=0.450\linewidth,keepaspectratio]{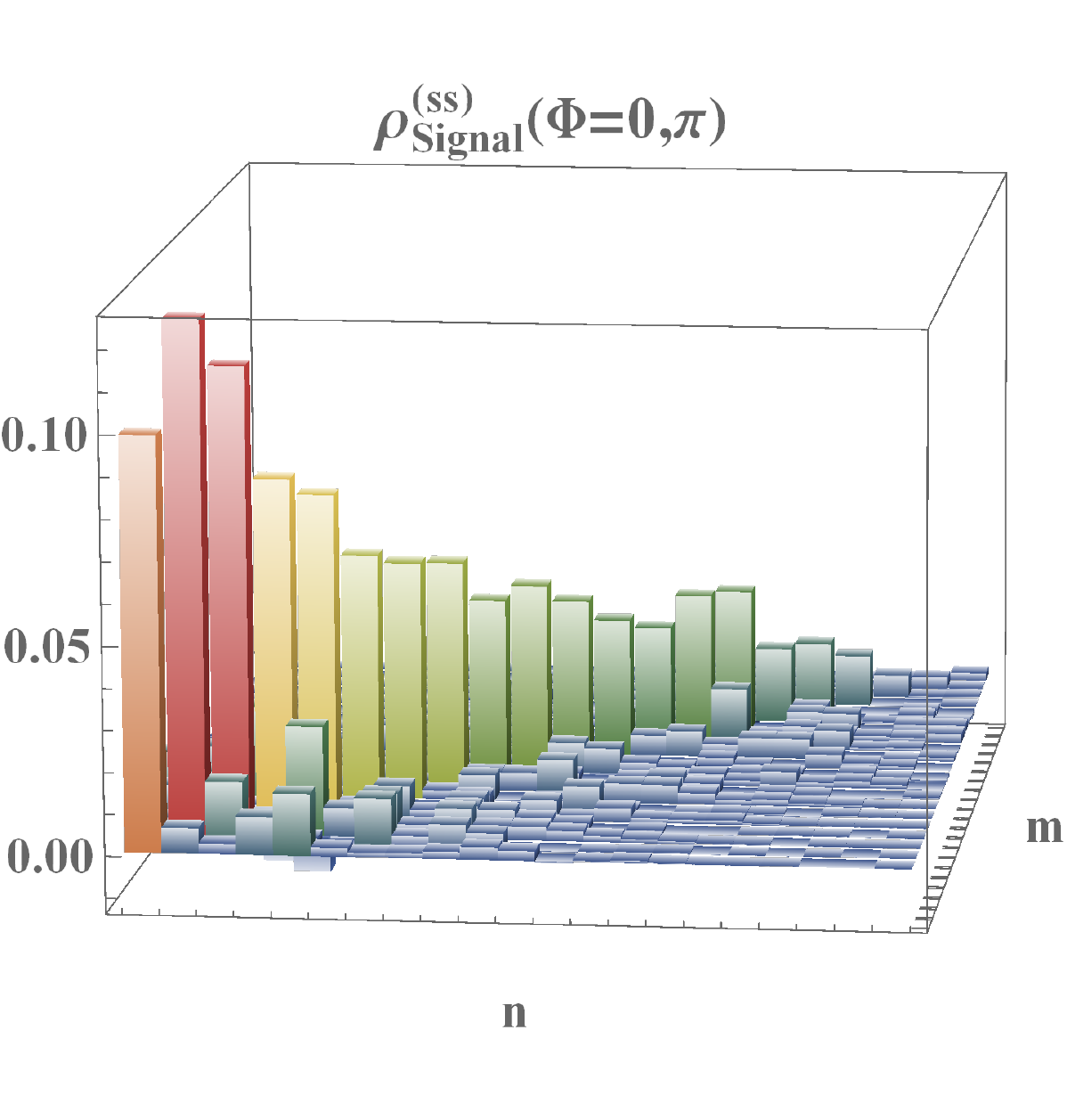}
		\label{fig:Num_DensitySignalPhi}}
	\caption{Steady state density matrix elements $\rho_{n,m}$ for the pump and signal(idler) modes, $\rho_\text{Pump}^{\left(ss\right)}\left(\Phi\right)$ and $\rho_\text{Signal}^{\left(ss\right)}\left(\Phi\right)$, respectively, for different values of the phase $\Phi$. Note that diagonal elements, $\rho_{n,n}$, denote the mode photon number distribution and that the steady state for the cases of $\Phi=0,\;\pi$ are the same.  For the signal(idler) modes, most of the distribution is captured when considering elements $\rho_{n,n}$ and $\rho_{n,n\pm1}$ for all cases of the phase, indicating the thermal-like nature of the steady state.}
	\label{fig:Num_Density} 
\end{figure*}

\noindent This is, up to a relative phase that can be absorbed into the coefficients, the steady state recursion relation for a constant pump.  In order to solve this, we make the approximation that $\sqrt{n_{s}n_{i}}\approx \sqrt{n_{s}\left(n_{s}+\Delta n\right)}$ where $\Delta n = n_{i}-n_{s}$ where we assume $\Delta n/n_{s}<<1$. This approximation is valid when considering that while the joint photon number distribution of the two-mode squeezed coherent states are centered around the diagonal line $n_{s}=n_{i}$, or $\Delta n=0$, most of the distribution is captured when considering $\Delta n = 0,\; \pm 1$. This is shown explicitly in Fig. \ref{fig:Num_Density}, where we plot the density matrices for the signal/pump modes for several values of the cumulative phase $\Phi$.  With this taken into account, the recursion relation becomes

\begin{equation}
		\left(\frac{n_{s}+n_{i}}{2}+1\right)C_{n_{s}+1,n_{i}+1}=\left(\frac{n_{s}+n_{i}}{2}\right)C_{n_{s}-1,n_{i}-1},
	\label{eqn:ss3}
\end{equation}  

\noindent with solutions

\begin{align}
	C_{n_{s},n_{i}} = &\left(\frac{\Gamma\left(\tfrac{1}{4}\left(n_{i}-n_{s}+4\right)\right)\Gamma\left(\tfrac{1}{4}\left(n_{i}-n_{s}+2\right)\right)}{\Gamma\left(\tfrac{1}{4}\left(n_{i}-n_{s}+2\right)\right)\Gamma\left(\tfrac{1}{4}\left(n_{i}+n_{s}+4\right)\right)}\right)\times \nonumber \\
	 &\times\Big[c_{1}\left(n_{i}-n_{s}\right)+\left(-1\right)^{n_{s}}c_{2}\left(n_{i}-n_{s}\right)\Big],
	\label{eqn:ss4}
\end{align}

\noindent where $c_{1}$ and $c_{2}$ are arbitrary functions dependent solely on the mode photon number difference $\Delta n$ and the choice of initial state. Note that the fact the solutions  are not factorisable can be taken as an indication of the entanglement between the signal and idler modes. To summarize, the reduced density matrix of a two mode squeezed state is thermal; for the state in question, the signal/idler modes are thermal-like at long times.  The discrepancy can be attributed to several things: first, unlike the two-mode squeezed state, the photon number correlations are not such that $\Delta n =0$. Diagonal elements are dominant, but off-diagonal terms exist.  Secondly, we are effectively ignoring the entanglement between signal/idler modes and the pump.         

We attempt one more approximation to illustrate the entanglement between the pump and signal/idler modes in the steady state for a non-constant pump.  Let us decompose the state in Eq. \ref{eqn:34} as

\begin{align}
	\ket{\psi\left(t\right)} = \sum_{n_{p_{0}}=0}^{\infty}&\sum_{\Delta n=-\infty}^{\infty}\sum_{n=\Delta n}^{n_{p_{0}}} C_{n_{p_{0}},\Delta n,n}\left(t\right) \times \nonumber \\
	&\times\ket{n_{p_{0}}-n}_{p}\ket{n+\Delta n}_{s}\ket{n-\Delta n}_{i},
	\label{eqn:Alsing1}
\end{align}

\noindent where $\Delta n=n_{i}-n_{s}$ is a constant of motion and we can view the sum in Eq. \ref{eqn:Alsing1} as a sum over each 'fixed' value of $n_{p_{0}}$ of a number of photons in the pump.  Note that the invariant operator $\hat{n}_{p}+\tfrac{1}{2}\left(\hat{n}_{s}+\hat{n}_{i}\right)$ yields $n_{p_{0}}$ for each basis state $\ket{n_{p_{0}}-n}_{p}\ket{n+\Delta n}_{s}\ket{n-\Delta n}_{i}$.  The Schr\"{o}dinger equation for the time-dependent state coefficients yields

\begin{widetext}
	\begin{equation}
		\dot{C}_{n_{p_{0}},\Delta n, n}=\sqrt{n_{p_{0}}-n}\sqrt{\left(n+1\right)+\Delta n}\sqrt{\left(n+1\right)-\Delta n}\;C_{n_{p_{0}},\Delta n, n+1} - \sqrt{\left(n_{p_{0}}+1\right)-n}\sqrt{n+\Delta n}\sqrt{n- \Delta n}\;C_{n_{p_{0}},\Delta n, n-1}.
		\label{eqn:Alsing2}
	\end{equation}
\end{widetext}

\noindent The point to note here is that all coefficients in Eq. \ref{eqn:Alsing2} can be considered to be at fixed $n_{p_{0}}$ and $\Delta n$, with only $n$ varying.  Thus, for a fixed number $n_{p_{0}}$ of pump photons, the Schr\"{o}dinger equation breaks up into a set of differential-difference equations in $n$ characterized by the constant difference $\Delta n$ between the signal and idler photon states.  At steady state, we have the equation

\begin{widetext}
	\begin{equation}
		\sqrt{n_{p_{0}}-n}\sqrt{\left(n+1\right)+\Delta n}\sqrt{\left(n+1\right)-\Delta n}\;C_{n_{p_{0}},\Delta n, n+1}=\sqrt{\left(n_{p_{0}}+1\right)-n}\sqrt{n+\Delta n}\sqrt{n- \Delta n}\;C_{n_{p_{0}},\Delta n, n-1},
		\label{eqn:Alsing3}
	\end{equation}
\end{widetext}

\noindent which is, in general, intractable algebraically.  To bring out the coupling between the pump and the signal/idler modes, we consider the regime where $n_{p_{0}}\gg n \gg \Delta n$.  This approximation allows us to expand the radicals to first order in $n_{p_{0}}$ and $\Delta n$ (dropping terms of order $\mathcal{O}\big[\left(n/n_{p_{0}}\right)^{2}\big]$ and $\mathcal{O}\big[\left(\Delta n/n\right)^{2}\big]$ and where terms of $\mathcal{O}\left[\left(\Delta n/n\right)\right]$ cancel) to obtain the following approximate equation as a function of $n$ for fixed $n_{p_{0}}$ and $\Delta n$

\begin{align}
	2\left(n_{p_{0}}+1\right)&\left(2n_{p_{0}}-n\right)\left(n+1\right)C_{n_{p_{0}},\Delta n,n+1} \approx \nonumber \\
	&\approx 2n_{p_{0}}\left(2\left(n_{p_{0}}+1\right)-n\right)nC_{n_{p_{0}},\Delta n,n-1},
	\label{eqn:Alsing4}
\end{align}

\noindent with solutions 

\begin{align}
	C_{n_{p_{0}},\Delta n,n} &= c_{1}\frac{\left(-1\right)^{n+1}\left(2n_{p_{0}}+1\right)\left(\tfrac{n_{p_{0}}}{n_{p_{0}}+1}\right)^{n/2}\Gamma\left(\tfrac{n+1}{2}\right)}{\sqrt{\pi}\left(n-2n_{p_{0}}-1\right)\Gamma\left(\tfrac{n+2}{2}\right)} \nonumber \\
	&- c_{2} \frac{\left(2n_{p_{0}}+1\right)\left(2n_{p_{0}}+1\right)\left(\tfrac{n_{p_{0}}}{n_{p_{0}}+1}\right)^{n/2}\Gamma\left(\tfrac{n+1}{2}\right)}{\sqrt{\pi}\left(n-2n_{p_{0}}-1\right)\Gamma\left(\tfrac{n+2}{2}\right)},
	\label{eqn:Alsing5}
\end{align}

\noindent with arbitrary constants $c_{1}$ and $c_{2}$.  Note that each term in Eq. \ref{eqn:Alsing5} nearly factorizes in $n_{p_{0}}$ (pump) and $n$ (signal/idler - recall that $\Delta n/n$ has canceled out to first order), except for the non-factorizable term $\left(n-2n_{p_{0}}-1\right)$ in the denominator.  It is this non-factorizability of $C_{n_{p_{0}},\Delta n,n}$ which indicates the (weakly) entangled nature between the pump and the signal/idler modes in the steady state in the examined regime $n_{p_{0}}\gg n\gg\Delta n$.

\section{\label{sec:level6} VI. Conclusion}

After reviewing the state statistics of coherently-stimulated parametric down-conversion with a constant pump field, we consider the case of a quantized pump field, initially taken to be a coherent state, with seeded coherent states in the signal and idler modes. For this case, we discuss how the state statistics evolve in time, with particular emphasis on the effects of the cumulative phase $\Phi=\theta_{s}+\theta_{i}-2\phi$, where $\theta_{s}$ and $\theta_{i}$ are the signal and idler coherent state phases, respectively, and $2\phi$ is the classical pump phase.  We include short-time analytic expressions obtained through perturbation for the state properties we discuss.  For long times, we employ a fourth order Runge-Kutta numerical integration method to find the state probability amplitudes and corresponding density matrix. Similar to the case of a constant pump, the statistics of the state rely solely on the cumulative phase value and not on the individual state phases. Furthermore, for transient times, we show that all three modes display sub-Poissonian statistics.  In addition to this, we discuss the entanglement properties of the state through calculation of the bipartite monotone, the logarithmic negativity. Using this, we consider the entanglement between signal/idler modes as well as between the signal/pump modes.  We go on to calculate the mutual information to directly find the degree of entanglement between the pump and joint signal-and-idler modes. We show that for long times, the steady state  displays thermal-like properties. This is shown by comparing the von Neumann entropy of the signal and pump modes with the entropy of an effective 'thermal' state of equal average photon number.  We close with a qualitative discussion of the two-mode signal/idler steady state valid for long times as well as a discussion on the entanglement between signal/idler modes for the regime $\Delta n/n_{s}$ and pump mode for the regime $n_{p_{0}} \gg n \gg \Delta n$.     

\section{\label{sec:level7} VII. Acknowledgements}

RJB acknowledges support from the NRC RAP.  CCG acknowledges support under AFRL Visiting Faculty Fellowship Program, AFRL Contract No. FA8750-116-3-6003.  PMA acknowledges support from the Air Force Office of Scientific Research.  Any opinions, findings and conclusions or recommendations expressed in this material are those of the author(s) and do not necessarily reflect the views of AFRL.

\end{document}